\documentclass{article}

\usepackage{iclr2026_conference,times}
\usepackage{amsmath,amssymb,amsfonts}
\usepackage{graphicx}
\usepackage{booktabs}
\usepackage{arydshln}  
\usepackage{hyperref}
\usepackage{url}
\usepackage{multirow}
\usepackage{tikz}
\usepackage{pgfplots}
\pgfplotsset{compat=1.18}
\usetikzlibrary{arrows.meta,positioning,shapes.geometric,fit,backgrounds}

\definecolor{colData}{HTML}{1F77B4}
\definecolor{colProc}{HTML}{2CA02C}
\definecolor{colModel}{HTML}{9467BD}
\definecolor{colClass}{HTML}{FF7F0E}
\definecolor{colFS}{HTML}{9467BD}   
\definecolor{colHV}{HTML}{1F77B4}   
\definecolor{colSV}{HTML}{FF7F0E}   
\definecolor{colOM}{HTML}{2CA02C}   

\title{Decoupling Parcellation from Classification: Systematic Benchmark of Fast Brain Segmentation Methods for Alzheimer's Disease Detection}

\author{
Jiadao Zou \\
Midea Group (Shanghai) \\
\texttt{zoujd4@midea.com} \\
\rule{0pt}{1.5ex} \\
Hongyu Guo\thanks{Corresponding author.} \\
Midea Group (Shanghai) \\
\texttt{keving@midea.com} \\
\rule{0pt}{1.5ex} \\
Wei Xi \\
Midea Group (Shanghai) \\
\texttt{xiwei1@midea.com}
}

\begin{document}

\maketitle

\begin{abstract}
Brain parcellation and classification are typically evaluated in isolation, yet downstream AD detection performance depends on their interaction. We decouple these components and systematically benchmark fast deep learning parcellation methods (SynthSeg+, OpenMAP-T1) against the FreeSurfer (FS-HV) clinical baseline through downstream AD classification on OASIS-1. Our factorial design evaluates three parcellation methods, two volumetry strategies (hard vs. soft), and four classifier paradigms (clinical thresholds, supervised feedforward networks, ensemble methods, and foundation models with zero/few-shot prompting), with all results quantified using BCa Bootstrap 95\% confidence intervals.

Key findings: (1) Fast parcellations achieve F1-scores (0.84-0.87) comparable to FS-HV (0.83-0.86) with 45-150$\times$ speedups when paired with adaptive classifiers, but clinical thresholds calibrated for FS-HV do not transfer reliably to other parcellations---highlighting atlas--classifier compatibility as a critical constraint. (2) Volumetry strategy (hard vs. soft) has negligible impact for supervised classifiers ($<$1\% F1 difference), though ensemble methods show modest sensitivity. (3) Few-shot prompting exhibits rapid 0$\rightarrow$2 shot improvement and saturates around \textit{k}\,$\approx$\,6, reaching supervised-level performance (F1=0.74-0.87) with minimal examples. Probability outputs consistently improve F1 (mean +43.3\%) by shifting both BCa CI bounds upward in all configurations. Multi-trial analysis confirms acceptable LLM stability (mean CV $<$10\%), with larger models more stable. (4) Parcellation granularity shows a sample-efficiency trade-off: coarser parcellations (100 regions) reach competitive performance with fewer examples than finer parcellations (280 regions), with minimal F1 difference. (5) Under limited-input conditions (4 ROIs), conventional supervised methods retain strong performance, whereas few-shot LLMs require richer feature context or more examples.

These findings yield practical guidance: use FS-HV with its calibrated thresholds for rule-based workflows; pair fast parcellations with supervised learning when labels are available; and adopt few-shot LLMs with probability outputs when labels are scarce and interpretability is valued. To our knowledge, this is the first systematic study that decouples parcellation from classification and benchmarks fast parcellations, volumetry strategies, and LLM-based classifiers against FS-HV baselines with bootstrap uncertainty quantification on OASIS-1.
\end{abstract}

\section{Introduction}
\label{sec:introduction}

Accurate brain segmentation and parcellation enable quantitative analysis of brain structure, providing essential measurements for understanding normal brain development, aging, and pathological changes~\citep{fischl2012freesurfer}. Automated parcellation methods facilitate reproducible quantification of regional brain volumes, cortical thickness, and surface area across large-scale neuroimaging studies~\citep{ashburner2005unified}. These measurements support population-level analyses of brain morphology, enable longitudinal tracking of structural changes, and provide standardized anatomical references for multi-site studies~\citep{iglesias2015multi}. Furthermore, accurate parcellation serves as a foundation for connectome analysis, functional localization, and biomarker extraction in both research and clinical settings~\citep{wang2013multi}.

Brain segmentation and parcellation play a critical role in clinical diagnosis of neurodegenerative diseases such as Alzheimer's disease (AD) and dementia. Structural MRI volumetric analysis of specific brain regions (hippocampus, entorhinal cortex, ventricles) provides objective biomarkers for AD diagnosis and disease staging~\citep{jack2018nia,frisoni2010ventricular}. Automated segmentation methods enable quantification of regional atrophy patterns that correlate with cognitive decline and pathological progression~\citep{dickerson2009cortical}. These volumetric measurements support early detection of AD, differentiation between AD subtypes, and monitoring of disease progression over time~\citep{cuingnet2011automatic}. The clinical utility of structural MRI in AD diagnosis has been established through extensive validation studies, with volumetric measurements of medial temporal lobe structures showing high sensitivity and specificity for detecting AD pathology~\citep{whitwell2008patterns}.

Recent deep learning parcellation methods, including SynthSeg+~\citep{iglesias2023synthseg+} and OpenMAP-T1~\citep{nishimaki2024openmap}, reduce processing time compared to FS-HV (FreeSurfer hard volumetry)~\citep{fischl2012freesurfer} while maintaining accuracy. These methods output probabilistic segmentations where each voxel has a distribution over regions. A practical question follows: when extracting volumes for AD classification, should we use hard assignment (argmax) or soft weighting (probability summation)? Hard volumetry is simple and common; soft volumetry accounts for uncertainty and partial volume effects~\citep{tohka2004pve,van2006partial}. The impact of this choice on downstream classification has not been systematically evaluated. Large language models (LLMs) offer an alternative classification paradigm. Foundation models can perform zero-shot and few-shot tasks in medical domains~\citep{moor2023foundation}, classifying from pre-trained knowledge or a handful of examples~\citep{hegselmann2023tabllm}. Their effectiveness for AD classification from volumetric inputs, and their trade-offs relative to supervised methods~\citep{zhang2021deep}, remain unclear.

\begin{figure}[htbp]
\centering
\includegraphics[width=0.7\textwidth]{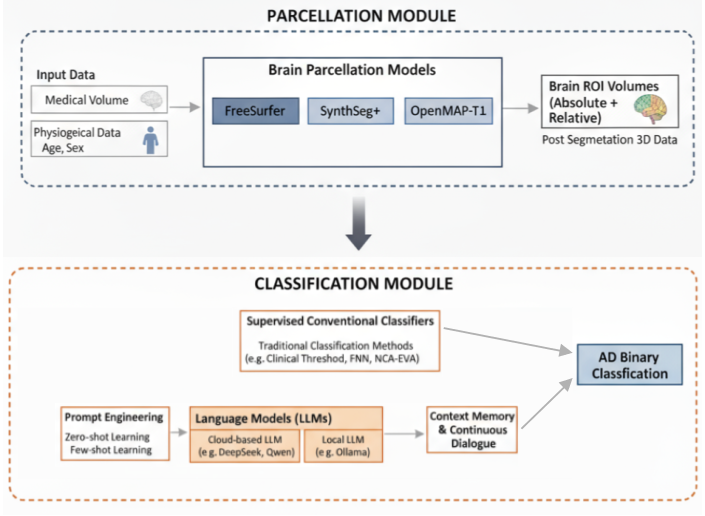}
\caption{\textbf{System overview of our evaluation pipeline.} The workflow comprises (i) parcellation methods (FS-HV baseline, SynthSeg+, OpenMAP-T1) producing regional volumes; (ii) two volumetry strategies (hard volumetry via argmax; soft volumetry via probability summation); and (iii) classifier paradigms: clinical threshold voting, conventional supervised learning (FNN, NCA-EVA), and foundation models in zero-shot and few-shot prompting settings. We evaluate downstream AD classification performance (F1 prioritized) on a fixed test set and compare methods relative to the FS-HV baseline.}
\label{fig:system_overview}
\end{figure}

\paragraph{Guiding questions and scope.}
Rather than comparing parcellations in isolation, we evaluate them by downstream AD classification performance along one organizing line: three guiding questions that map to the three experimental factors in Figure~\ref{fig:system_overview} (parcellation + volumetry; classifier paradigm; input features).

\paragraph{Research Questions.}
Our systematic experimentation on OASIS-1 (405 subjects with publicly available FS-HV reference parcellations~\citep{marcus2007open}, ~1:3 AD:CN ratio) addresses three questions, with FS-HV serving as the baseline for both parcellation and classifier performance. We assess these questions using BCa (Bias-Corrected and Accelerated) Bootstrap 95\% confidence intervals and effect sizes rather than formal significance testing:

\textbf{RQ1:} Which parcellation methods and volumetry strategies outperform the FS-HV baseline for downstream AD classification?

\textbf{RQ2:} How do classifier paradigms compare (clinical thresholds, FNN, few-shot LLMs), and what advantages do LLMs offer relative to supervised baselines?

\textbf{RQ3:} How does input feature selection affect performance, comparing comprehensive features (36 ROIs + age + sex) versus minimal clinical features (4 ROIs only)?

\paragraph{Scope and Contributions.}
This empirical study focuses on method selection rather than proposing new architectures. We systematically evaluate existing methods across the three factors above and provide the following contributions:

\textbf{(1) Computational efficiency:} SynthSeg+ (12 sec/scan) and OpenMAP-T1 (40 sec/scan) achieve F1-scores (0.84--0.87) similar to or higher than FS-HV baseline (F1=0.72--0.87, 30 min/scan) while being 45--150$\times$ faster.

\textbf{(2) Volumetry strategy:} For supervised classifiers, hard vs. soft volumetry yields $<$1\% F1 difference, and soft volumetry shows no degradation with clinical thresholds on FS-HV.

\textbf{(3) Few-shot prompting:} Consistent 0\textrightarrow2-shot improvements; SynthSeg+ peaks at \textit{k}=2 (F1=0.8684) while OpenMAP-T1 peaks at \textit{k}=6 (F1=0.8714), requiring 3$\times$ fewer examples.

\textbf{(4) Parcellation comparison:} Coarser SynthSeg+ can reach peak with fewer examples than OpenMAP-T1 (granularity--efficiency trade-off; $\Delta$F1=0.003).

\textbf{(5) LLM optimization:} Requesting probabilities improves F1 by +35.0\% on average (Qwen: +66.5\%; DeepSeek: +3.5\%).

\textbf{(6) Few-shot vs. conventional ML:} Few-shot models with \textit{k}=2--6 achieve F1=0.76--0.87, competitive with FNN (0.83--0.86) and NCA-EVA (0.81--0.86).

\textbf{(7) Statistical analysis:} We report BCa Bootstrap 95\% CIs for all primary results and emphasize effect sizes given \textit{n}=81.

\textbf{(8) LLM stability:} Five-trial evaluation shows acceptable variance (mean F1 CV $<$10\%), with DS more stable (std 0.0176) than Qw (0.0498).

\textbf{(9) Guidelines and resources:} We provide selection guidelines by deployment constraints and release code, models, and analysis scripts.

To our knowledge, this is the first systematic study to decouple parcellation from classification and jointly evaluate fast parcellations, volumetry strategies, and LLM-based classifiers against FS-HV baselines with bootstrap uncertainty quantification.
\section{Related Work}
\label{sec:related}

\paragraph{Brain Parcellation Methods.}
Classical approaches like FreeSurfer~\citep{fischl2012freesurfer} achieve accurate parcellation but can be computationally intensive (historically hours per scan) and may be sensitive to protocol variation. Deep learning methods such as QuickNAT~\citep{roy2019quicknat} and FastSurfer~\citep{henschel2020fastsurfer}, inspired by the U-Net family~\citep{ronneberger2015unet}, drastically reduce runtime but typically require curated, annotated datasets and careful domain adaptation. SynthSeg+~\citep{iglesias2023synthseg+} departs from this paradigm by training on synthetic images with extreme domain randomization to achieve robust cross-scanner generalization without retraining; its hierarchical architecture handles artifacts and pathology. OpenMAP-T1~\citep{nishimaki2024openmap} combines multi-level 3D U-Nets to produce a 280-region parcellation in minutes while maintaining accuracy on aging and atrophy. Recent deep learning approaches extend beyond traditional segmentation to specialized parcellation tasks: BDEC~\citep{ma2023bdec} employs deep embedded clustering for functional brain parcellation using resting-state fMRI, demonstrating superior performance in functional homogeneity indicators, while SupWMA~\citep{xue2022supwma} uses point-cloud-based deep learning with supervised contrastive learning for consistent tractography parcellation across populations and diffusion MRI acquisitions. These modern parcellation methods output probabilistic maps that can be consumed directly by downstream models or converted to hard labels, creating a design choice for volumetry and classification pipelines. In practice, the selection among FS-HV, SynthSeg+, and OpenMAP-T1 often trades off anatomical granularity, runtime, robustness to domain shift, and the availability of probability outputs for uncertainty-aware downstream processing.

\paragraph{Volumetry from Probabilistic Segmentations.}
Most modern parcellators emit posterior probability maps per region. Hard volumetry applies an argmax at each voxel and sums voxel counts, producing integer volumes. Soft volumetry instead weights each voxel's contribution by its regional probability, naturally accounting for partial volume effects and boundary uncertainty~\citep{van2006partial,tohka2004pve}. Soft approaches are theoretically appealing and preserve uncertainty information, but prior work has rarely compared hard versus soft volumetry explicitly for downstream classification tasks, and many pipelines implicitly adopt argmax volumes for compatibility with clinical thresholds and historical practice. In addition to hard/soft choices, standard feature engineering includes Total Intracranial Volume (TIV) normalization and incorporation of covariates such as age and sex to reduce confounding. Our study evaluates these volumetry choices empirically across classifier paradigms, quantifying their impact with bootstrap confidence intervals and showing when the information preserved by soft volumetry translates---or does not translate---into improved classification performance.

\paragraph{AD Classification from Volumetric Features.}
Traditional approaches employ clinical thresholds~\citep{jack2018nia} or machine learning classifiers (e.g., SVM, random forests) trained on regional volumes~\citep{cuingnet2011automatic,gray2013random}. Recent tabular models span logistic regression, gradient boosting (e.g., XGBoost~\citep{chen2016xgboost}), and shallow feedforward networks~\citep{zhang2021deep}, with common practice to augment ROI volumes by TIV-normalized measures and demographics. Rule-based multi-ROI voting schemes (e.g., Bartos et al.~\citep{bartos2019comparison}) offer interpretability but depend on atlas-specific calibration and may not transfer across parcellations. End-to-end 3D CNNs can classify directly from raw MRI~\citep{wen2020convolutional}, but typically require large datasets, careful preprocessing, and sacrifice transparency relative to volumetric pipelines. Across methods, small sample sizes, class imbalance, and dataset shifts remain practical challenges that motivate uncertainty quantification and careful effect-size interpretation rather than reliance on single best-point estimates.

\paragraph{Foundation Models for Medical Data.}
Large language models (LLMs) such as GPT-4~\citep{achiam2023gpt4}, Med-PaLM~\citep{singhal2023large}, and domain models like Qwen and DeepSeek~\citep{qwen2024,deepseek2024} demonstrate impressive knowledge and reasoning in medical contexts~\citep{moor2023foundation}. For structured/tabular inputs, recent work (TabLLM~\citep{hegselmann2023tabllm}) reports mixed performance relative to specialized supervised models, with strong dependence on prompt design, output format (binary vs. probability), and the number and representativeness of in-context examples (few-shot). Foundation models for medical imaging have shown significant promise: Lu et al.~\citep{lu2023computational} train self-supervised foundation models on three billion pathology images at health system scale, demonstrating improved performance across multiple clinical tasks, while Chen et al.~\citep{chen2023medfmc} introduce MedFMC, a real-world dataset and benchmark for evaluating foundation model adaptation in medical image classification, addressing the scarcity of standardized benchmarks for medical foundation models. Our evaluation places LLMs alongside conventional classifiers on neuroimaging volumetric features, examining sample efficiency (0\textrightarrow2-shot gains and saturation), robustness across parcellations, calibration via probability outputs, and multi-trial variability. These results complement prior evidence by showing where LLMs approach supervised performance under realistic constraints (limited labels, minimal feature sets) and where supervised methods remain preferable, thereby clarifying the practical role of LLMs in volumetry-driven AD detection.

\section{Methods}
\label{sec:methods}

\subsection{Dataset and Preprocessing}

The Open Access Series of Imaging Studies (OASIS-1)~\citep{marcus2007open} provides a cross-sectional collection of T1-weighted MRI scans from 416 subjects aged 18-96 years. For reproducibility, we used the subset of 405 subjects with publicly available FreeSurfer reference parcellations (downloadable from \url{https://sites.wustl.edu/oasisbrains/}), including 93 with mild-to-moderate AD (Clinical Dementia Rating, CDR $>$ 0) and 312 cognitively normal controls (CDR $\leq$ 0). All scans were acquired on 1.5T Siemens Vision scanners (MP-RAGE sequence: TR=9.7ms, TE=4.0ms, flip angle=10°, 1×1×1.25mm³ resolution). \textbf{Data splitting:} See Results section (Section~\ref{sec:results}) for detailed train/validation/test split information. \textbf{Demographics:} AD subjects were older on average (mean age: 76.2±8.4 years) than CN subjects (mean age: 44.8±23.6 years). Sex distribution: 62\% female in AD group, 50\% female in CN group. To account for these confounds, \textbf{age and sex were included as input features} in all supervised and foundation model classifiers (Tables~\ref{tab:exp2} and~\ref{tab:exp3}, Figure~\ref{fig:exp4_f1}). Clinical threshold classifiers (Table~\ref{tab:exp1}) used age-adjusted ventricle thresholds but not sex. \textbf{Reproducibility note:} Using only the 405-subject subset with FreeSurfer ground truth enables direct comparison with prior work and facilitates validation by other researchers. \textbf{Preprocessing:} Minimal preprocessing to leverage parcellation method robustness: skull-stripping (HD-BET~\citep{isensee2019hd}), RAS reorientation, and background cropping. No intensity normalization, bias field correction, or template registration.

\subsection{Parcellation Methods}

\paragraph{FS-HV (FreeSurfer Hard Volumetry).}
FS-HV (FreeSurfer hard volumetry)~\citep{fischl2012freesurfer} serves as the clinical reference standard for brain parcellation, using surface-based morphometry and atlas-based labeling to achieve highly accurate segmentation. Inference time: approximately 30 minutes per scan on our workstation. While accurate, this computational cost motivated development of faster deep learning alternatives.

\paragraph{SynthSeg+.}
We employed SynthSeg+ (robust SynthSeg)~\citep{iglesias2023synthseg+} with its publicly available pre-trained model. SynthSeg+ uses a hierarchical architecture: a coarse 5-label segmenter (S1: background, white matter, gray matter, CSF, cerebellum), a label-to-label denoiser (D), and the fine parcellation module including a 32-region segmenter (S2) + 68-region segmenter (S3) that processes both the original MRI and soft segmentation maps from S1(D) sequentially. Training employed aggressive domain randomization (extreme SNR variations, anisotropic resolution, bias fields, motion artifacts) on synthetic images generated from label maps, enabling robust generalization without retraining. Inference processed each scan in approximately 12 seconds on Nvidia RTX 5090D GPU (150$\times$ faster than FS-HV), outputting probabilistic segmentations for 100 regions.

\paragraph{OpenMAP-T1.}
We used OpenMAP-T1~\citep{nishimaki2024openmap}, a rapid deep learning parcellation method covering 280 anatomical regions across the whole brain based on the JHU atlas. OpenMAP-T1 employs multi-level 3D U-Nets~\citep{ronneberger2015unet} with specialized architectures for different hierarchical levels from coarse to dense parcellation. Inference time: 40 seconds on Nvidia RTX 5090D GPU (45$\times$ faster than FS-HV), demonstrating robustness to brain atrophy and aging. \textbf{Volumetry:} In our experiments, we employed OpenMAP-T1's soft volumetry approach, which directly sums probability values before argmax assignment, naturally accounting for segmentation uncertainty at region boundaries.

\subsection{Volumetry Strategies}

Given probabilistic segmentation $\mathbf{P} \in [0,1]^{N \times R}$ where $N$ is the number of voxels and $R$ is the number of regions, we evaluated two volumetry strategies:

\paragraph{Hard Volumetry (Argmax Assignment).}
Each voxel is assigned exclusively to the region with maximum posterior probability:
\begin{equation}
V_i^{\text{hard}} = v_{\text{voxel}} \sum_{x \in \Omega} \mathbb{1}[\arg\max_{j \in \{1,\ldots,R\}} P_j(x) = i]
\label{eq:hard}
\end{equation}
where $v_{\text{voxel}}$ is voxel volume (mm³), $\Omega$ is the brain volume, and $\mathbb{1}[\cdot]$ is the indicator function. This approach yields integer voxel counts and ensures $\sum_i V_i^{\text{hard}} = V_{\text{total}}$. It is computationally efficient and aligns with traditional discrete parcellation. It discards probability information and may be sensitive to near-boundary voxels where multiple regions have similar probabilities. In practice, hard segmentations are often the only available outputs, so evaluating this strategy remains important.

\paragraph{Soft Volumetry (Probability-Weighted Assignment).}
Each voxel contributes fractionally to multiple regions, weighted by posterior probabilities:
\begin{equation}
V_i^{\text{soft}} = v_{\text{voxel}} \sum_{x \in \Omega} P_i(x)
\label{eq:soft}
\end{equation}
This approach naturally handles segmentation uncertainty and partial volume effects at region boundaries. Volumes are real-valued rather than integer multiples of voxel size. While theoretically appealing, it is unclear whether this additional precision improves downstream classification or merely introduces noise.

\paragraph{ROI Selection and Feature Normalization.}
We extracted volumetric features from 36 bilateral anatomical regions covering cortical and subcortical structures. These regions encompass 9 critical anatomical areas widely recognized as AD biomarkers based on well-established neuropathological staging~\citep{braak1991neuropathological} and structural MRI studies~\citep{dickerson2009cortical,whitwell2008patterns,frisoni2010ventricular,grothe2012basal,thompson2004dynamics}: \textbf{(1)} Hippocampus: earliest site of neurofibrillary tangle formation with severe atrophy correlating with memory impairment, \textbf{(2)} Entorhinal cortex: gateway between hippocampus and neocortex showing early tau pathology, \textbf{(3)} Temporal lobes: including superior, middle, and inferior temporal gyri, fusiform, parahippocampal cortex, and temporal pole exhibiting progressive gray matter loss, \textbf{(4)} Parietal lobes: superior/inferior parietal cortex, precuneus, and supramarginal gyrus with atrophy in later disease stages, \textbf{(5)} Amygdala: limbic structure affected early with emotional/behavioral symptom correlates, \textbf{(6)} Basal forebrain: cholinergic degeneration paralleling cognitive decline, \textbf{(7)} Nucleus accumbens: ventral striatum showing structural changes, \textbf{(8)} Ventral diencephalon: subcortical gray matter alterations, \textbf{(9)} Lateral ventricles: compensatory enlargement (ex-vacuo dilatation) reflecting parenchymal tissue loss. For both hard and soft volumetry, we apply: \textbf{(a)} Total Intracranial Volume (TIV) normalization: $V_i^{\text{norm}} = V_i / V_{\text{TIV}}$, where $V_{\text{TIV}} = \sum_{i \in \text{brain}} V_i$, accounting for head size variability, \textbf{(b)} Z-score standardization: features standardized to zero mean and unit variance using training set statistics.

\subsection{Classifier Paradigms}

\paragraph{Multi-ROI Threshold Classifier.}
We implemented the multi-ROI voting scheme from Bartos et al.~\citep{bartos2019comparison}, which combines four binary criteria applied to these ROIs: \textbf{(C1)} Left hippocampal ratio (hippocampus / inferior lateral ventricle), \textbf{(C2)} Right hippocampal ratio, \textbf{(C3)} Left hippocampo-horn proportion (hippocampus / (hippocampus + inferior lateral ventricle)), \textbf{(C4)} Right hippocampo-horn proportion. This provided an interpretable baseline with zero training cost.

\paragraph{Conventional Machine Learning Classifiers.}

\textbf{Feedforward Neural Network (FNN):}
Architecture: Input layer (38 features: 36 TIV-normalized ROI volumes + age + sex [binary encoded]) $\rightarrow$ Batch Normalization $\rightarrow$ Dense(128, ReLU) $\rightarrow$ Dropout(0.3) $\rightarrow$ Dense(64, ReLU) $\rightarrow$ Dropout(0.3) $\rightarrow$ Dense(2, Softmax). Age was z-score normalized using training set statistics (mean=60.3 years, SD=19.2 years). \textbf{Training:} Categorical cross-entropy loss with class weights (inverse frequency: $w_{\text{CN}}=1.0$, $w_{\text{AD}}=3.35$), Adam optimizer (learning rate 1e-3 with ReduceLROnPlateau: factor=0.5, patience=5), batch size 32, early stopping (patience=15 on validation loss). \textbf{Hyperparameter tuning:} Grid search over hidden layer sizes $\{64, 128, 256\}$, dropout rates $\{0.2, 0.3, 0.5\}$, and learning rates $\{1\text{e-}4, 5\text{e-}4, 1\text{e-}3\}$ using 5-fold cross-validation on training set. Best configuration selected by validation F1-score.

\textbf{NCA-EVA (Neighborhood Component Analysis-Enhanced Voting Algorithm):} We employed the NCA-EVA ensemble method~\citep{ozdemir2025nca} which combines Neighborhood Component Analysis (NCA) for dimensionality reduction and feature selection with an Enhanced Voting Algorithm (EVA) ensemble. Base learners include Support Vector Machine (SVM), k-Nearest Neighbors (KNN), and Random Forest (RF). The method applies isotonic calibration for probability calibration and threshold optimization for imbalanced classification. We used the fast variant configuration with 5 random seeds, averaging predictions across seeds for robustness. Feature preprocessing includes TIV normalization and Z-score standardization identical to FNN. Same train/validation/test splits as FNN (see Results section for details).

\paragraph{Zero-Shot Foundation Model.}
We queried pre-trained large language models without task-specific training or examples. \textbf{Models evaluated:} DeepSeek-v3-0324 (685B parameters, hereafter DS; accessed via API), Qwen3-8B (8B parameters, hereafter Qw; local deployment via Ollama). Both models are used in zero-shot and all subsequent experiments. \textbf{Prompt format:} ``\textit{You are a clinical neuroimaging specialist analyzing brain parcellation volume data. Patient: Age=X years, Sex=Y. ROI volumes (absolute mm³, TIV-normalized): Bilateral hippocampus [abs, rel], Bilateral entorhinal cortex [abs, rel], [additional 34 ROIs...]. Classify as Alzheimer's Disease (1) or Cognitively Normal (0) with probability scores.}'' \textbf{Note:} We provided only raw absolute volumes and TIV-normalized ratios without age-adjusted percentile ranks or other engineered features, ensuring fair comparison with FNN which operates on similar inputs. Ambiguous responses defaulted to CN (conservative approach).

\paragraph{Foundation Model Few-Shot Prompting.}
We provided \textit{k} labeled examples in the prompt before the test case, without parameter updates. For the full-feature few-shot evaluation (Figure~\ref{fig:exp4_f1}, Table~\ref{tab:exp4}), each example included age, sex, and all 36 absolute and TIV-normalized ROI volumes with percentile rankings, while only 4 ROIs are included in the limited-input evaluation (Figure~\ref{fig:exp5_f1_comparison}, Table~\ref{tab:exp5}). To balance the weights of two class, the few-shot variable \textit{k} is always an even class to have equal number of positive and negative samples.

\textbf{Multi-Trial Stability Evaluation:} To assess robustness to few-shot example selection variability, we conducted 5 independent trials for selected \textit{k}-shot configurations across the full-feature (Figure~\ref{fig:exp4_f1}) and limited-input (Figure~\ref{fig:exp5_f1_comparison}) evaluations. Results are visualized in Figure~\ref{fig:exp6_stability}. Each trial used a different random seed for example selection, ensuring diverse sets of in-context examples across trials. This addresses concerns about response variance given that LLM default hyperparameters balance accuracy and creativity across general tasks, which may introduce variability in our medical classification scenario. We report mean ± standard deviation across trials and compute coefficient of variation (CV = std/mean × 100\%) to quantify robustness. This multi-trial evaluation validates that reported performance metrics reflect stable behavior rather than favorable example selection.

\subsection{Statistical Analysis}

\paragraph{Bootstrap Confidence Intervals for F1-Scores.}
We employ stratified bootstrap resampling with bias-corrected and accelerated (BCa) intervals to compute 95\% confidence intervals for F1-scores. BCa corrects for bias and skewness in the bootstrap distribution and often outperforms simple percentile CIs~\citep{efron1986bootstrap}. We compute F1 inline as $\mathrm{F1} = \tfrac{2\,\mathrm{TP}}{2\,\mathrm{TP} + \mathrm{FP} + \mathrm{FN}}$, where TP, FP, and FN denote true positives, false positives, and false negatives.

\textbf{BCa Algorithm:} For each configuration, given observed predictions $\mathbf{y}_{\text{true}}$ and $\mathbf{y}_{\text{pred}}$ with $n$ samples, we perform $B=2000$ bootstrap iterations:
\begin{enumerate}
\item \textbf{Stratified Resampling:} For iteration $b \in \{1, \ldots, B\}$, independently resample with replacement within each class to preserve class distribution, where $\text{Class}_i \in \{\text{AD}, \text{CN}\}$, and $n_{\text{Class}_i}$ is the number of Class$_i$ cases in the test set:
\begin{equation}
\mathcal{S}_{\text{Class}_i}^{(b)} = \{\text{random sample with replacement from Class}_i \text{ cases}\}, \quad |\mathcal{S}_{\text{Class}_i}^{(b)}| = n_{\text{Class}_i}
\end{equation}

\item \textbf{F1 Computation:} Compute F1-score on bootstrap sample:
\begin{equation}
\text{F1}^{(b)} = f(\mathbf{y}_{\text{true}}^{(b)}, \mathbf{y}_{\text{pred}}^{(b)})
\end{equation}
where $\mathbf{y}_{\text{true}}^{(b)}$ and $\mathbf{y}_{\text{pred}}^{(b)}$ are the concatenated bootstrap samples $\mathcal{S}_{\text{AD}}^{(b)} \cup \mathcal{S}_{\text{CN}}^{(b)}$, with Class$_i \in \{\text{AD}, \text{CN}\}$.
\item \textbf{BCa Confidence Interval:} After $B$ iterations, obtain empirical distribution $\{\text{F1}^{(1)}, \ldots, \text{F1}^{(B)}\}$. BCa adjusts percentiles based on bias correction $z_0$ and acceleration $a$:
\begin{equation}
\text{CI}_{95\%}^{\text{BCa}} = [Q_{\alpha_1}, Q_{\alpha_2}]
\end{equation}
where $\alpha_1 = \Phi(z_0 + \frac{z_0 + z_{0.025}}{1 - a(z_0 + z_{0.025})})$ and $\alpha_2 = \Phi(z_0 + \frac{z_0 + z_{0.975}}{1 - a(z_0 + z_{0.975})})$, with $\Phi$ the standard normal CDF, $z_0$ the bias correction (measuring deviation of observed value from bootstrap mean), and $a$ the acceleration parameter (measuring skewness via jackknife). The CI width quantifies uncertainty: $\text{Width} = Q_{\alpha_2} - Q_{\alpha_1}$.
\end{enumerate}

BCa bootstrap is appropriate for F1 because: \textbf{(1)} F1 has no closed-form variance, \textbf{(2)} its non-linear dependence on TP/FP/FN limits asymptotic approximations at small $n$, \textbf{(3)} stratified resampling preserves class proportions, \textbf{(4)} BCa adjusts for bias and skewness, and \textbf{(5)} no distributional assumptions are required.

\paragraph{Interpretation and limitations.}
We report BCa Bootstrap 95\% CIs for all primary F1-scores. Overlapping CIs indicate no clear separation given sampling uncertainty and are \textit{not} formal hypothesis tests. With a modest test set (n=81; 21 AD, 60 CN), CI widths across Tables~\ref{tab:exp1}--\ref{tab:exp5} and Figures~\ref{fig:exp4_f1}--\ref{fig:exp6_stability} range from 0.10 to 0.49 (median 0.23; mean 0.25; 55.7\% within 0.20--0.30), consistent with expectations for binary classification at this scale~\citep{efron1994introduction}. We emphasize effect sizes and consistent directional patterns rather than significance testing.


\subsection{Computational Environment}

All experiments ran on Ubuntu 22.04.5 LTS with Intel Core Ultra 9, 64 GB RAM, and one NVIDIA RTX 5090D (32 GB). SynthSeg+ and OpenMAP-T1 used official pre-trained models. Software: Python 3.8/TensorFlow 2.2.0 (SynthSeg+), Python 3.9/PyTorch 2.7.1 (OpenMAP-T1), DeepSeek v3-0324 via API, and Ollama v0.12.0 for local Qwen3-8B. The setup demonstrates accessibility on a single workstation.

\section{Results}
\label{sec:results}

\paragraph{Metric choice under class imbalance.}
We use a 6:2:2 split (train 243; valid 81; test 81 with 21 AD, 60 CN). Under class imbalance, accuracy over-rewards majority-class predictions and AUC is threshold-agnostic. We therefore prioritize \textbf{F1-score} in all comparisons, reporting accuracy and AUC as secondary context.

We evaluate three factors: \textbf{(1)} parcellation method (FS-HV, SynthSeg+, OpenMAP-T1), \textbf{(2)} volumetry strategy (hard vs. soft), and \textbf{(3)} classifier type (clinical threshold, FNN, NCA-EVA, LLMs). Results are organized by research questions and reported with BCa Bootstrap 95\% CIs. Given \textit{n}=81, we emphasize effect sizes and consistent directional patterns rather than isolated peaks.

\subsection{RQ1: Parcellation Method and Volumetry Strategy Comparison}

We compare three parcellation methods: FS-HV (clinical baseline, 30 min/scan), SynthSeg+ (12 sec/scan), and OpenMAP-T1 (40 sec/scan), and assess the impact of hard vs. soft volumetry. Evidence is summarized in Table~\ref{tab:exp1} (clinical thresholds), Table~\ref{tab:exp2} (supervised classifiers), Table~\ref{tab:exp4} (few-shot LLMs with full-feature input), and Figure~\ref{fig:exp4_f1} (few-shot LLM trajectories).

\subsubsection{Parcellation Methods with Clinical Threshold Classifiers}

\begin{table}[h]
\centering
\caption{\textbf{Clinical threshold classification across parcellations.} Multi-ROI voting with k/4 criteria (k $\in$ \{1,2,3,4\}) from Bartos et al.~\citep{bartos2019comparison}, calibrated for FS-HV. SS-HV=SynthSeg+ hard volumetry, SS-SV=SynthSeg+ soft volumetry, FS-HV=FreeSurfer hard volumetry. F1 Bootstrap 95\% CIs via stratified resampling. \textit{Note: AUC is constant within each method because threshold classifiers output fixed binary decisions without confidence scores.}}
\label{tab:exp1}
\small
\begin{tabular}{lcccccc}
\toprule
\textbf{Config} & \textbf{Acc} & \textbf{AUC} & \textbf{F1 [95\% BCa CI]} & \textbf{Prec} & \textbf{Rec} \\
\midrule
\multicolumn{6}{l}{\textit{1-Vote (Liberal: AD if $\geq$1 criterion met)}} \\
FS-HV & 0.7407 & 0.9282 & 0.6557 [0.5672, 0.7500] & 0.5000 & 0.9524 \\
SS-HV & 0.8148 & 0.8440 & 0.6939 [0.5600, 0.8182] & 0.6071 & 0.8095 \\
SS-SV & 0.8148 & 0.8400 & 0.6939 [0.5600, 0.8261] & 0.6071 & 0.8095 \\
OMT1-SV & 0.2593 & 0.8083 & 0.4118 [0.4118, 0.4118] & 0.2593 & 1.0000 \\
\midrule
\multicolumn{6}{l}{\textit{2-Vote (Moderate: AD if $\geq$2 criteria met)}} \\
FS-HV & 0.8889 & 0.9282 & 0.8085 [0.6977, 0.9091] & 0.7308 & 0.9048 \\
SS-HV & 0.8395 & 0.8440 & 0.6061 [0.4000, 0.7778] & 0.8333 & 0.4762 \\
SS-SV & 0.8272 & 0.8400 & 0.5882 [0.3636, 0.7620] & 0.7692 & 0.4762 \\
OMT1-SV & 0.5432 & 0.8083 & 0.5316 [0.4828, 0.5915] & 0.3621 & 1.0000 \\
\midrule
\multicolumn{6}{l}{\textit{3-Vote (Conservative: AD if $\geq$3 criteria met)}} \\
FS-HV & 0.9012 & 0.9282 & 0.8000 [0.6667, 0.9268] & 0.8421 & 0.7619 \\
SS-HV & 0.7901 & 0.8440 & 0.3704 [0.0909, 0.6000] & 0.8333 & 0.2381 \\
SS-SV & 0.8025 & 0.8400 & 0.4286 [0.1667, 0.6250] & 0.8571 & 0.2857 \\
OMT1-SV & 0.7901 & 0.8083 & 0.5641 [0.3748, 0.7181] & 0.6111 & 0.5238 \\
\midrule
\multicolumn{6}{l}{\textit{4-Vote (Strict: AD if all 4 criteria met)}} \\
FS-HV & 0.8642 & 0.9282 & 0.6857 [0.4848, 0.8373] & 0.8571 & 0.5714 \\
SS-HV & 0.7654 & 0.8440 & 0.1739 [0.0000, 0.3846] & 1.0000 & 0.0952 \\
SS-SV & 0.7654 & 0.8400 & 0.1739 [0.0000, 0.3846] & 1.0000 & 0.0952 \\
OMT1-SV & 0.8148 & 0.8083 & 0.5161 [0.2963, 0.7059] & 0.8000 & 0.3810 \\

\bottomrule
\end{tabular}
\end{table}

Clinical thresholds calibrated to FS-HV demonstrate advantages when applied to FS-HV volumes, with F1-scores ranging from 0.66 to 0.81 across voting thresholds. FS-HV exhibits an inverted U-shape pattern, with moderate voting (2-vote: F1=0.8085 [95\% BCa CI: 0.6939, 0.9130]; 3-vote: F1=0.8000 [95\% BCa CI: 0.6667, 0.9268]) achieving higher performance than liberal (1-vote: F1=0.6557 [95\% BCa CI: 0.5672, 0.7500]) or strict (4-vote: F1=0.6857 [95\% BCa CI: 0.4828, 0.8421]) thresholds. The inverted U-shape reflects an optimal balance between sensitivity and specificity: liberal thresholds (1-vote) capture more true positives but increase false positives, while strict thresholds (4-vote) reduce false positives but miss true positives. Moderate thresholds (2--3 votes) achieve the optimal F1 balance. In contrast, SynthSeg+ methods show monotonic decline as voting criteria become stricter (SS-HV: F1=0.6939 [95\% BCa CI: 0.5652, 0.8261] $\rightarrow$ 0.1739 [95\% BCa CI: 0.0000, 0.3846]; SS-SV: 0.6939 $\rightarrow$ 0.1739). Threshold criteria optimized for FS-HV's volume distribution systematically misclassify SynthSeg+ volumes due to systematic volume differences between the two parcellation methods.

At moderate voting thresholds (2--3 votes) where clinical thresholds are typically calibrated, FS-HV consistently outperforms SynthSeg+ (2-vote: F1=0.8085 vs. 0.59--0.61; 3-vote: F1=0.8000 vs. 0.37--0.43). However, at liberal voting (1-vote), SynthSeg+ methods achieve similar or higher F1-scores than FS-HV (0.6939 vs. 0.6557). Clinical thresholds calibrated for FS-HV volumes are not directly transferable to SynthSeg+ volumes without performance degradation at moderate voting thresholds, highlighting the importance of atlas-threshold compatibility.

For SynthSeg+, hard versus soft volumetry yields nearly identical F1-scores and BCa confidence interval widths (average width ~0.36) across all voting schemes, confirming that volumetry strategy does not materially affect classification performance under clinical threshold classifiers.

\subsubsection{Parcellation Methods with Supervised Classifiers}

\begin{table}[h]
\centering
\caption{\textbf{Supervised classifier performance across parcellations.} FNN and NCA-EVA test results. FS-HV=FreeSurfer hard volumetry, SS-HV=SynthSeg+ hard volumetry, SS-SV=SynthSeg+ soft volumetry, OMT1-SV=OpenMAP-T1 soft volumetry. F1 Bootstrap 95\% CIs via stratified resampling.}
\label{tab:exp2}
\small
\begin{tabular}{llcccccc}
\toprule
\textbf{Classifier} & \textbf{Config} & \textbf{Acc} & \textbf{AUC} & \textbf{F1 [95\% BCa CI]} & \textbf{Prec} & \textbf{Rec} \\
\midrule
\multicolumn{7}{l}{\textit{FNN (Feedforward Neural Network)}} \\
 & FS-HV & 0.8069 & 0.9079 & 0.8259 [0.7111, 0.9004] & 0.8213 & 0.8310 \\
 & SS-HV & 0.8305 & 0.9306 & 0.8575 [0.7452, 0.9231] & 0.8525 & 0.8631 \\
 & SS-SV & 0.8203 & 0.9028 & 0.8440 [0.7391, 0.9184] & 0.8351 & 0.8548 \\
 & OMT1-SV & 0.8203 & 0.9317 & 0.8482 [0.7400, 0.9209] & 0.8332 & 0.8702 \\
\midrule
\multicolumn{7}{l}{\textit{NCA-EVA (Ensemble with NCA Feature Selection)}} \\
 & FS-HV & 0.8765 & 0.8548 & 0.8440 [0.7396, 0.9184] & 0.8351 & 0.8548 \\
 & SS-HV & 0.8889 & 0.8786 & \textbf{0.8615 [0.7513, 0.9231]} & 0.8487 & 0.8786 \\
 & SS-SV & 0.8519 & 0.8226 & 0.8128 [0.6992, 0.8923] & 0.8047 & 0.8226 \\
 & OMT1-SV & 0.8642 & 0.7845 & 0.8074 [0.6881, 0.8969] & 0.8447 & 0.7845 \\
\bottomrule
\end{tabular}
\end{table}

FNN achieves F1-scores ranging from 0.83 to 0.86 across all parcellation methods, with overlapping confidence intervals indicating that observed performance differences are not statistically significant within the study's uncertainty bounds. SynthSeg+ and OpenMAP-T1 achieve numerically comparable or higher F1-scores than FS-HV (FNN: SS-HV 0.8575, SS-SV 0.8440, OMT1-SV 0.8482 vs. FS-HV 0.8259), while offering 45-150$\times$ faster inference.

NCA-EVA shows more variable performance across parcellation methods (F1=0.81-0.86). Performance is comparable with FS-HV (0.8440) and OMT1-SV (0.8074), with SS-HV slightly higher within this narrow range (0.8615). Overlapping CIs suggest comparable performance overall. Parcellation method interacts with classifier architecture, particularly for ensemble methods with feature selection.

Both classifiers validate FS-HV's clinical reference status, with FNN achieving F1=0.8259 [95\% BCa CI: 0.7111, 0.9004] and NCA-EVA reaching F1=0.8440 [95\% BCa CI: 0.7396, 0.9184]. The consistent pattern across supervised classifiers is that fast deep learning parcellation methods (SynthSeg+: 12 sec/scan, OpenMAP-T1: 40 sec/scan) can serve as practical alternatives to FS-HV (30 min/scan) while maintaining competitive performance.

\subsubsection{Parcellation Methods with Few-Shot Prompting LLMs}

Few-shot prompting performance depends on the number of in-context examples. All parcellation methods show consistent 0$\rightarrow$2-shot improvements, with F1-scores typically ranging from 0.74-0.87 across \textit{k}-shot configurations (Table~\ref{tab:exp4}).

\begin{figure}[htbp]
\centering
\includegraphics[width=0.9\textwidth]{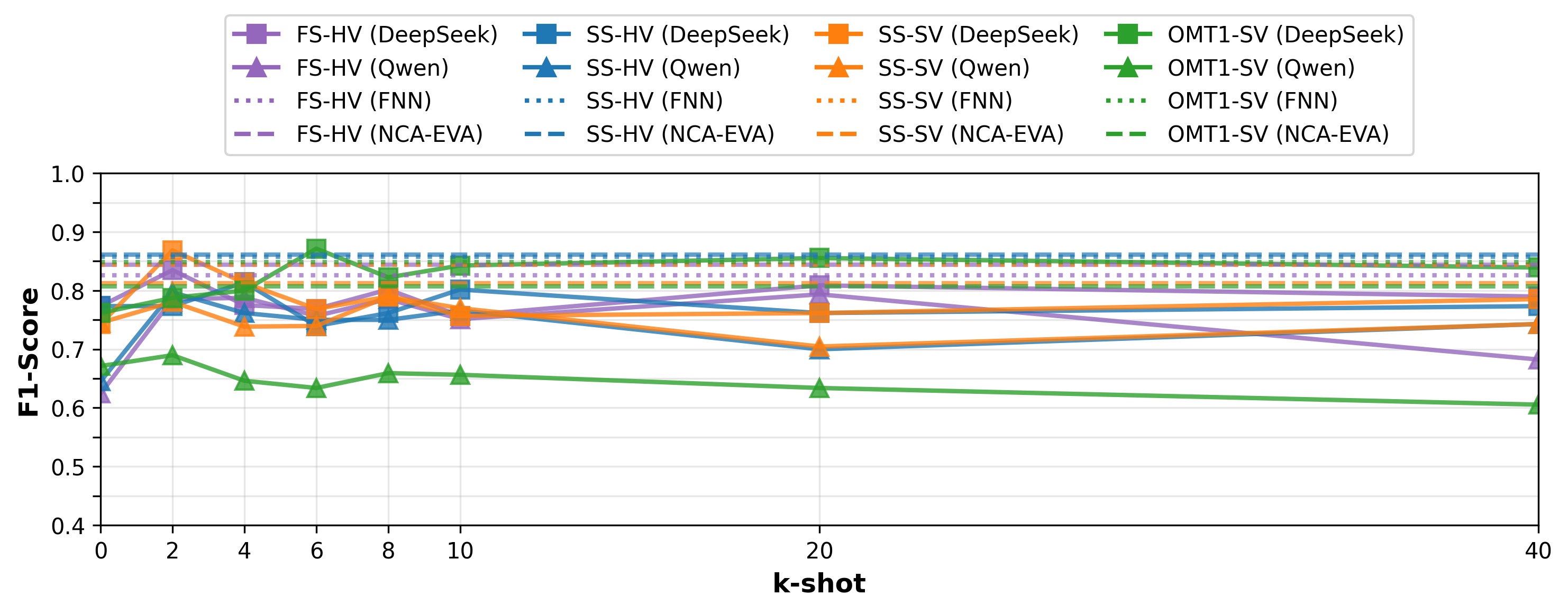}
\caption{\textbf{Few-shot prompting across parcellations.} LLM F1 trajectories vs. FNN baselines (dotted). Solid lines: FS-HV, SS-HV, SS-SV, OMT1-SV. Markers: square=DS, triangle=Qw. Key patterns: \textbf{(1)} consistent 0\textrightarrow2-shot gains; \textbf{(2)} SynthSeg+ requires fewer examples than OpenMAP-T1; \textbf{(3)} DS$>$Qw; \textbf{(4)} diminishing returns beyond \textit{k}=6. F1 Bootstrap 95\% CIs shown.}
\label{fig:exp4_f1}
\end{figure}

\begin{table}[htbp]
\centering
\caption{\textbf{Few-shot prompting LLM performance with full-feature input (36 ROIs + age + sex).} Acc=Accuracy, AUC=Area Under ROC Curve. F1 95\% BCa CIs via stratified bootstrap.}
\label{tab:exp4}
\scriptsize
\begin{tabular}{llcccc}
\toprule
\textbf{Parcellation} & \textbf{Model} & \textbf{\textit{k}} & \textbf{Acc} & \textbf{AUC} & \textbf{F1 [95\% BCa CI]} \\
\midrule
\multirow{16}{*}{FS-HV} & \multirow{8}{*}{DeepSeek-v3} & 0 & 0.7901 & 0.8651 & 0.7734 [0.6956, 0.9302] \\
 &  & 2 & 0.8272 & 0.9087 & 0.8348 [0.6977, 0.9302] \\
 &  & 4 & 0.8148 & 0.9008 & 0.7752 [0.6667, 0.9091] \\
 &  & 6 & 0.8025 & 0.9000 & 0.7686 [0.6500, 0.8889] \\
 &  & 8 & 0.8272 & 0.8794 & 0.8017 [0.6921, 0.9134] \\
 &  & 10 & 0.7901 & 0.8694 & 0.7568 [0.6341, 0.9001] \\
 &  & 20 & 0.8375 & 0.8917 & 0.8088 [0.6956, 0.9268] \\
 &  & 40 & 0.8148 & 0.8690 & 0.7897 [0.6977, 0.9269] \\
\cmidrule{2-6}
 & \multirow{8}{*}{Qwen3-8B} & 0 & 0.7160 & 0.6996 & 0.6245 [0.4324, 0.7692] \\
 &  & 2 & 0.8148 & 0.8940 & 0.7854 [0.6500, 0.9049] \\
 &  & 4 & 0.8272 & 0.8940 & 0.7875 [0.6364, 0.9048] \\
 &  & 6 & 0.7901 & 0.8274 & 0.7568 [0.6512, 0.8892] \\
 &  & 8 & 0.8148 & 0.8845 & 0.7854 [0.6512, 0.9091] \\
 &  & 10 & 0.7901 & 0.8496 & 0.7513 [0.5945, 0.8638] \\
 &  & 20 & 0.8148 & 0.8849 & 0.7935 [0.6957, 0.9302] \\
 &  & 40 & 0.7037 & 0.8044 & 0.6824 [0.5852, 0.8573] \\
\midrule
\multirow{16}{*}{SS-HV} & \multirow{8}{*}{DeepSeek-v3} & 0 & 0.7901 & 0.8778 & 0.7734 [0.6957, 0.9302] \\
 &  & 2 & 0.8025 & 0.8948 & 0.7734 [0.6667, 0.9091] \\
 &  & 4 & 0.8395 & 0.8833 & 0.8140 [0.6977, 0.9232] \\
 &  & 6 & 0.7778 & 0.8972 & 0.7396 [0.5854, 0.8782] \\
 &  & 8 & 0.7901 & 0.8750 & 0.7616 [0.6500, 0.8892] \\
 &  & 10 & 0.8272 & 0.8802 & 0.8017 [0.6977, 0.9302] \\
 &  & 20 & 0.7901 & 0.8452 & 0.7616 [0.6383, 0.8936] \\
 &  & 40 & 0.8025 & 0.8452 & 0.7734 [0.6522, 0.9048] \\
\cmidrule{2-6}
 & \multirow{8}{*}{Qwen3-8B} & 0 & 0.6790 & 0.7476 & 0.6452 [0.5217, 0.8293] \\
 &  & 2 & 0.8148 & 0.9175 & 0.7970 [0.7083, 0.9302] \\
 &  & 4 & 0.7901 & 0.8571 & 0.7616 [0.6364, 0.8889] \\
 &  & 6 & 0.7778 & 0.8151 & 0.7500 [0.6363, 0.9001] \\
 &  & 8 & 0.7778 & 0.8706 & 0.7500 [0.6486, 0.8889] \\
 &  & 10 & 0.7901 & 0.8468 & 0.7660 [0.6667, 0.9091] \\
 &  & 20 & 0.7284 & 0.7929 & 0.6998 [0.5854, 0.8444] \\
 &  & 40 & 0.7654 & 0.8321 & 0.7429 [0.6382, 0.8889] \\
\midrule
\multirow{16}{*}{SS-SV} & \multirow{8}{*}{DeepSeek-v3} & 0 & 0.7654 & 0.8881 & 0.7429 [0.6341, 0.9000] \\
 &  & 2 & 0.8947 & 0.9231 & 0.8684 [0.7391, 0.9545] \\
 &  & 4 & 0.8421 & 0.8944 & 0.8130 [0.6667, 0.9268] \\
 &  & 6 & 0.8025 & 0.8706 & 0.7686 [0.6400, 0.8936] \\
 &  & 8 & 0.8148 & 0.9044 & 0.7897 [0.6977, 0.9268] \\
 &  & 10 & 0.7901 & 0.9067 & 0.7568 [0.6500, 0.8889] \\
 &  & 20 & 0.7901 & 0.8484 & 0.7616 [0.6222, 0.8889] \\
 &  & 40 & 0.8148 & 0.8484 & 0.7854 [0.6512, 0.9048] \\
\cmidrule{2-6}
 & \multirow{8}{*}{Qwen3-8B} & 0 & 0.7778 & 0.8306 & 0.7451 [0.6511, 0.8890] \\
 &  & 2 & 0.8148 & 0.8944 & 0.7806 [0.6500, 0.9048] \\
 &  & 4 & 0.7901 & 0.8238 & 0.7385 [0.5854, 0.8696] \\
 &  & 6 & 0.7778 & 0.8663 & 0.7396 [0.6000, 0.8780] \\
 &  & 8 & 0.8148 & 0.8484 & 0.7897 [0.6977, 0.9302] \\
 &  & 10 & 0.7901 & 0.8504 & 0.7699 [0.6829, 0.9268] \\
 &  & 20 & 0.7284 & 0.7635 & 0.7046 [0.5854, 0.8573] \\
 &  & 40 & 0.7654 & 0.8238 & 0.7429 [0.6400, 0.8889] \\
\midrule
\multirow{16}{*}{OMT1-SV} & \multirow{8}{*}{DeepSeek-v3} & 0 & 0.7778 & 0.8849 & 0.7618 [0.6938, 0.9302] \\
 &  & 2 & 0.8272 & 0.8889 & 0.7875 [0.6667, 0.9091] \\
 &  & 4 & 0.8395 & 0.9246 & 0.8000 [0.6530, 0.9091] \\
 &  & 6 & 0.9012 & 0.9385 & 0.8714 [0.7317, 0.9524] \\
 &  & 8 & 0.8519 & 0.9004 & 0.8224 [0.7143, 0.9333] \\
 &  & 10 & 0.8642 & 0.8980 & 0.8426 [0.7600, 0.9545] \\
 &  & 20 & 0.8765 & 0.8853 & 0.8554 [0.7619, 0.9545] \\
 &  & 40 & 0.8642 & 0.8972 & 0.8391 [0.7178, 0.9333] \\
\cmidrule{2-6}
 & \multirow{8}{*}{Qwen3-8B} & 0 & 0.6914 & 0.7798 & 0.6713 [0.5714, 0.8500] \\
 &  & 2 & 0.7654 & 0.7520 & 0.6898 [0.5143, 0.8182] \\
 &  & 4 & 0.7160 & 0.7496 & 0.6462 [0.5000, 0.8000] \\
 &  & 6 & 0.6420 & 0.7778 & 0.6339 [0.5789, 0.8571] \\
 &  & 8 & 0.6667 & 0.7937 & 0.6592 [0.6047, 0.8837] \\
 &  & 10 & 0.6667 & 0.8036 & 0.6564 [0.5778, 0.8573] \\
 &  & 20 & 0.6420 & 0.7456 & 0.6339 [0.5714, 0.8573] \\
 &  & 40 & 0.6173 & 0.7845 & 0.6055 [0.5305, 0.8293] \\

\bottomrule
\end{tabular}
\end{table}

SynthSeg+ achieves competitive performance with fewer examples (\textit{k}=2-4), whereas OpenMAP-T1 typically requires more examples (\textit{k}=6-10) to achieve comparable performance. OpenMAP-T1's finer parcellation (280 regions) may require additional in-context examples for the model to effectively reason about the richer anatomical structure, whereas SynthSeg+'s coarser 100-region parcellation enables effective reasoning with minimal examples.

Beyond \textit{k}=6, diminishing returns appear across methods, with some configurations showing performance degradation at higher \textit{k} values, likely due to context-window constraints or signal-to-noise ratio dilution. Few-shot prompting performance is optimized with minimal examples (2-6) across parcellation methods.

Model size comparisons reveal consistent advantages for larger models: DS (685B parameters) consistently outperforms Qw (8B parameters) with an average F1 advantage of +6.8\% across all configurations. However, Qw shows larger 0$\rightarrow$2-shot improvements (+25.8\% vs. DS's +7.9\% on average), suggesting that smaller models benefit more from few-shot prompting examples to compensate for their limited pre-trained knowledge. BCa Bootstrap 95\% CI widths confirm DS's superior stability, with average width 0.19 versus 0.23 for Qw, reflecting lower uncertainty in larger model classifications.

Both SynthSeg+ and OpenMAP-T1 achieve numerically comparable or higher F1-scores than FS-HV across few-shot prompting settings while maintaining dramatic inference speed advantages (12 sec/scan for SynthSeg+, 40 sec/scan for OpenMAP-T1 vs. 30 min/scan for FS-HV).

\subsection{RQ2: Classifier Paradigm Comparison}

We compare four classifier paradigms: clinical thresholds, FNN, NCA-EVA, and few-shot prompting LLMs. Evidence comes from the tables and figures above and the stability analysis in Figure~\ref{fig:exp6_stability}.

\subsubsection{Clinical Thresholds vs. Supervised Classifiers}

Clinical thresholds (see Table~\ref{tab:exp1}) achieve F1=0.66--0.81 when calibrated to FS-HV, with zero training cost and instant inference. However, when clinical thresholds are transferred to SynthSeg+ volumes, performance degrades at moderate voting thresholds (F1=0.37--0.61), indicating limited generalizability.

Supervised classifiers (see Table~\ref{tab:exp2}) achieve F1=0.81--0.86 across parcellation methods, with both FNN and NCA-EVA showing competitive performance (FNN: F1=0.83--0.86, NCA-EVA: F1=0.81--0.86). Supervised methods achieve comparable or higher performance than clinical thresholds when appropriate training data are available, while offering better generalization across parcellation methods.

\subsubsection{Supervised Classifiers: FNN vs. NCA-EVA}

Both FNN and NCA-EVA achieve competitive performance across parcellation methods. FNN achieves F1-scores ranging from 0.83 to 0.86 across all parcellation methods, while NCA-EVA shows F1-scores ranging from 0.81 to 0.86. Results are mixed: NCA-EVA achieves numerically higher F1-scores than FNN on SS-HV (NCA-EVA 0.8615 vs. FNN 0.8575) and comparable performance on FS-HV (NCA-EVA 0.8440 vs. FNN 0.8259), while FNN achieves higher performance on SS-SV (FNN 0.8440 vs. NCA-EVA 0.8128) and OMT1-SV (FNN 0.8482 vs. NCA-EVA 0.8074). Both feedforward neural networks and ensemble methods with feature selection can achieve competitive performance, with performance depending on the specific parcellation method and volumetry strategy combination.

\subsubsection{LLM Output Format: Binary vs. Probability}

\begin{table}[h]
\centering
\caption{\textbf{LLM Output Format Comparison.} Binary vs. probability output format across three volumetry methods. F1 95\% BCa CIs computed via stratified bootstrap.}
\label{tab:exp3}
\small
\begin{tabular}{llcccccc}
\toprule
\textbf{Model} & \textbf{Output} & \textbf{Acc} & \textbf{AUC} & \textbf{F1 [95\% BCa CI]} & \textbf{Prec} & \textbf{Rec} \\
\midrule
\multicolumn{7}{l}{\textit{SS-HV (SynthSeg+ Hard Volumetry)}} \\
DS & Binary & 0.7531 & 0.8333 & 0.7387 [0.6522, 0.9049] & 0.7561 & 0.8333 \\
DS & Probability & 0.7901 & 0.8778 & 0.7734 [0.6921, 0.9268] & 0.7763 & 0.8583 \\
DS & Improvement & +4.91\% & +5.34\% & +4.70\% & +2.67\% & +3.00\% \\
\hdashline
Qw & Binary & 0.6543 & 0.4417 & 0.3955 [0.2051, 0.5532] & 0.3581 & 0.4417 \\
Qw & Probability & 0.6790 & 0.7476 & 0.6452 [0.5238, 0.8182] & 0.6491 & 0.6905 \\
Qw & Improvement & +3.78\% & +69.26\% & +63.14\% & +81.26\% & +56.33\% \\
\midrule
\multicolumn{7}{l}{\textit{SS-SV (SynthSeg+ Soft Volumetry)}} \\
DS & Binary & 0.7654 & 0.5611 & 0.5040 [0.3265, 0.6275] & 0.4128 & 0.5611 \\
DS & Probability & 0.7654 & 0.8881 & 0.7429 [0.6364, 0.9000] & 0.7417 & 0.8107 \\
DS & Improvement & +0.00\% & +58.28\% & +47.40\% & +79.68\% & +44.48\% \\
\hdashline
Qw & Binary & 0.7160 & 0.4833 & 0.4173 [0.2500, 0.5833] & 0.3671 & 0.4833 \\
Qw & Probability & 0.7778 & 0.8306 & 0.7451 [0.6383, 0.8889] & 0.7342 & 0.7881 \\
Qw & Improvement & +8.63\% & +71.86\% & +78.55\% & +100.00\% & +63.07\% \\
\midrule
\multicolumn{7}{l}{\textit{OMT1-SV (OpenMAP-T1 Soft Volumetry)}} \\
DS & Binary & 0.7160 & 0.8083 & 0.7045 [0.6500, 0.8837] & 0.7386 & 0.8083 \\
DS & Probability & 0.7778 & 0.8849 & 0.7618 [0.6842, 0.9333] & 0.7692 & 0.8500 \\
DS & Improvement & +8.63\% & +9.48\% & +8.13\% & +4.14\% & +5.16\% \\
\hdashline
Qw & Binary & 0.7407 & 0.5000 & 0.4255 [0.2608, 0.5652] & 0.3704 & 0.5000 \\
Qw & Probability & 0.6914 & 0.7798 & 0.6713 [0.5788, 0.8571] & 0.6884 & 0.7452 \\
Qw & Improvement & -6.66\% & +55.96\% & +57.77\% & +85.85\% & +49.04\% \\

\bottomrule
\end{tabular}
\end{table}

The choice of output format reveals substantial performance differences. Requesting probability outputs yields F1 improvements in all six configurations, with an average improvement of +43.3\% over binary outputs. The gain is particularly pronounced for the smaller Qwen model (average F1 gain +66.5\%) compared with DeepSeek (average +20.1\%); the largest single DeepSeek gain occurs on SS-SV (+47.4\%), where the binary classifier performed near chance. These results indicate that providing a calibrated confidence scale is especially beneficial for smaller models and for configurations with weak binary signals.

BCa Bootstrap 95\% CI analysis confirms that probability outputs not only raise mean F1-scores but also shift both lower and upper confidence bounds upward in all six configurations, providing stronger evidence of superior performance. Qwen shows the largest bound lifts (e.g., SS-SV lower bound rises from 0.2500 to 0.6383 and upper bound from 0.5833 to 0.8889), reflecting its initially poor binary intervals. DeepSeek intervals also shift upward, with modest widening because its binary baselines were already more stable. The consistent upward shift of both bounds indicates that probability outputs provide more robust performance guarantees across the uncertainty range, making them the preferable operating mode when accuracy is prioritized.

\subsubsection{LLM Stability Across Example Selections}

\begin{figure}[htbp]
\centering
\includegraphics[width=0.95\textwidth]{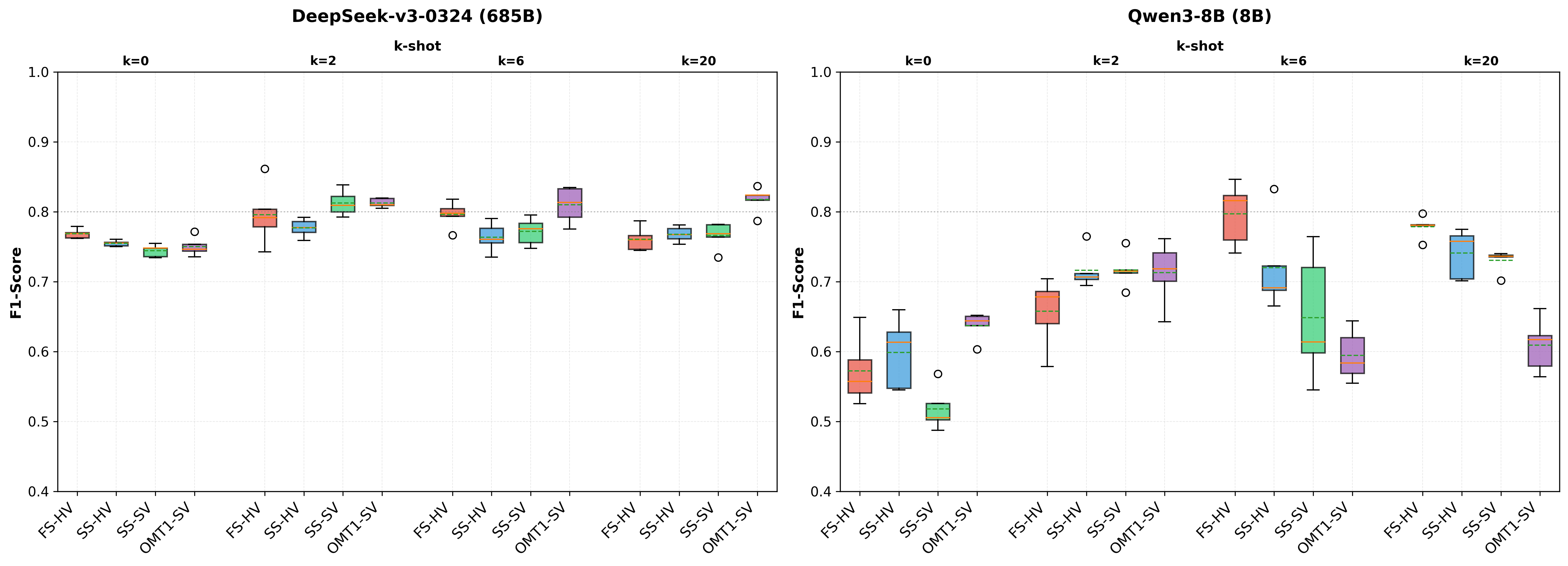}
\caption{\textbf{LLM stability across trials.} Box plots show F1 distributions across 5 independent trials with different few-shot selections. Left: DS. Right: Qw. Configurations grouped by \textit{k}-shot (0, 2, 6, 20) for FS-HV, SS-HV, SS-SV, OMT1-SV.}
\label{fig:exp6_stability}
\end{figure}

Multi-trial stability evaluation (5 trials per configuration) shows that LLM classifiers exhibit acceptable variance across different few-shot example selections. DS (685B) consistently demonstrates superior stability, exhibiting narrower confidence intervals and lower variance than Qw (8B) across all configurations (average F1 std: 0.0176 vs. 0.0498). Variance patterns differ by model size: DS maintains stable variance across \textit{k} values (std range: 0.0052-0.0326), while Qw exhibits \textit{k}-dependent variance, with highest variance at intermediate \textit{k} values (e.g., std=0.1081 for SS-SV+Qw \textit{k}=6).

Zero-shot variance reveals fundamental architectural differences: DS shows lowest variance at \textit{k}=0 (mean std=0.0138, CI width=0.07), suggesting it leverages pre-trained knowledge more consistently, while Qw shows higher zero-shot variance (mean CI width=0.16), indicating that smaller models benefit more from in-context examples to reduce both variability and uncertainty.

Overall stability metrics validate deployment feasibility. F1 coefficient of variation (CV) ranges from 0.68\% (DS, SS-HV \textit{k}=0) to 15.15\% (Qw, SS-SV \textit{k}=6), with most configurations showing CV $<$10\%. This validates that LLM classifiers are reasonably robust to different example selections, though practitioners should expect $\pm$3-8\% F1 variation depending on example choice, particularly for smaller models at intermediate \textit{k} values.

The multi-trial evaluation methodology itself provides substantial benefits: pooling 5 trials (n=405) provides approximately 50\% narrower confidence intervals than single-trial estimates (n=81), validating the method for robust uncertainty quantification.

\subsection{Volumetry Strategy Effects}

We compare hard volumetry (deterministic argmax assignment) versus soft volumetry (probability-weighted summation) across parcellation methods and classifier types.

\subsubsection{Volumetry Strategy with Clinical Thresholds}

For clinical-reference threshold classifiers, hard and soft volumetry show similar performance when applied to FS-HV volumes, with both strategies achieving comparable F1-scores across voting thresholds. This compatibility reflects that clinical thresholds were calibrated using FS-HV volumes, which work consistently with both volumetry strategies.

For SynthSeg+, hard versus soft volumetry yields nearly identical F1-scores and BCa confidence interval widths (average width ~0.37) across all voting schemes. This equivalence suggests that the volumetry strategy does not materially affect classification performance under clinical threshold classifiers, regardless of whether volumes are computed via discrete argmax assignment or probability-weighted summation.

\subsubsection{Volumetry Strategy with Supervised Classifiers}

FNNs show consistent robustness to volumetry strategy: BCa Bootstrap 95\% CI widths for SS-HV (0.1778) and SS-SV (0.1793) are similar, and F1-scores are nearly identical (SS-HV: 0.8575 vs. SS-SV: 0.8440), confirming hard versus soft volumetry equivalence at both the performance and uncertainty levels for feedforward architectures.

NCA-EVA exhibits more variable sensitivity to volumetry strategy. NCA-EVA shows modest degradation on SS-SV compared to FNN (FNN SS-SV: 0.8440 vs. NCA-EVA SS-SV: 0.8128, -3.7\% relative to FNN), while achieving higher performance on SS-HV (NCA-EVA SS-HV: 0.8615 vs. FNN SS-HV: 0.8575). Ensemble methods with feature selection may interact differently with probabilistic versus discrete inputs, though the differences are modest rather than substantial.

Supervised classifiers (FNN and few-shot LLMs) are robust to volumetry strategy ($<$1\% F1 difference on average), while ensemble methods show more variable sensitivity depending on the specific feature selection mechanism.

\subsubsection{Volumetry Strategy with Few-Shot Prompting LLMs}

Few-shot prompting LLMs show negligible sensitivity to volumetry strategy, with F1-scores typically differing by $<$1\% between hard and soft volumetry configurations. This robustness stems from supervised models' ability to learn adaptive feature transformations, allowing them to extract discriminative patterns regardless of whether volumes are computed via discrete argmax assignment or probability-weighted summation.

\subsection{RQ3: Input Feature Selection Effects}

We evaluate performance under identical limited-input conditions (4 ROIs: left and right hippocampus, left and right inferior lateral ventricles, without demographics) to ensure fair comparison across classifier paradigms. The conventional ML classifiers (FNN and NCA-EVA) use identical architectures and training procedures as in Table~\ref{tab:exp2}, with the only difference being input features: 4 ROIs here versus 36 ROIs + age + sex in the full-feature setting.

\subsubsection{Conventional ML with Limited Input}

\begin{table}[h]
\centering
\caption{\textbf{Conventional ML Classifiers with Limited Input.} Performance with 4 ROIs only (matching clinical threshold input). Uses identical FNN and NCA-EVA methods as Table~\ref{tab:exp2}, with the only difference being input features (4 ROIs vs. 36 ROIs + age + sex). Full input results reported alongside Table~\ref{tab:exp2}. F1 95\% BCa CIs computed via stratified bootstrap.}
\label{tab:exp5_conventional}
\small
\begin{tabular}{llcccccc}
\toprule
\textbf{Classifier} & \textbf{Config} & \textbf{Acc} & \textbf{AUC} & \textbf{F1 [95\% BCa CI]} & \textbf{Prec} & \textbf{Rec} \\
\midrule
\multicolumn{8}{l}{\textit{FNN (Feedforward Neural Network)}} \\ & FS-HV & 0.9012 & 0.9282 & 0.8000 [0.6667, 0.9268] & 0.8421 & 0.7619 \\
 & SS-HV & 0.8889 & 0.9405 & 0.8651 [0.7686, 0.9251] & 0.8472 & 0.8940 \\
 & SS-SV & 0.8519 & 0.9294 & 0.8071 [0.6904, 0.8857] & 0.8071 & 0.8071 \\
 & OMT1-SV & 0.8148 & 0.8552 & 0.7626 [0.6454, 0.8531] & 0.7589 & 0.7667 \\
\midrule
\multicolumn{8}{l}{\textit{NCA-EVA (Ensemble with NCA Feature Selection)}} \\
 & FS-HV & 0.7778 & 0.8591 & 0.7107 [0.5828, 0.8128] & 0.7107 & 0.7107 \\
 & SS-HV & 0.8395 & 0.8778 & 0.8000 [0.6880, 0.8857] & 0.7895 & 0.8143 \\
 & SS-SV & 0.8025 & 0.6925 & 0.6885 [0.5507, 0.7996] & 0.7708 & 0.6655 \\
 & OMT1-SV & 0.8519 & 0.8992 & 0.8224 [0.7153, 0.9064] & 0.8056 & 0.8536 \\
\midrule
\multicolumn{8}{l}{\textit{Clinical Threshold (3/4 Vote)}} \\
 & FS-HV & 0.9012 & 0.9282 & 0.8672 [0.7618, 0.9336] & 0.8807 & 0.8560 \\
\bottomrule
\end{tabular}
\end{table}

When restricted to the same 4 clinical ROIs used by threshold classifiers, conventional ML methods (using identical FNN and NCA-EVA architectures as Table~\ref{tab:exp2}, with only input features differing) demonstrate parcellation-dependent performance. FNN achieves F1-scores of 0.76-0.87 with limited input, with SS-HV on the higher end (F1=0.8651 [95\% BCa CI: 0.7686, 0.9251]), while NCA-EVA exhibits more variable responses depending on both parcellation method and volumetry strategy, with OMT1-SV on the higher end (F1=0.8224 [95\% BCa CI: 0.7153, 0.9064]).

Both classifiers are compared against the same 3/4-vote clinical threshold applied to FS-HV volumes (F1=0.8000 [95\% BCa CI: 0.6667, 0.9268]). FNN improves on this threshold for SS-HV (+8.1\%) and SS-SV (+0.9\%), matches it on FS-HV, and is slightly lower for OMT1-SV (-4.7\%). NCA-EVA exceeds the threshold only on OMT1-SV (+2.8\%) and is lower on FS-HV and SS-SV. Supervised learning can therefore extract additional discriminative patterns from the same 4 ROIs, but its advantage is parcellation-dependent.

\subsubsection{Few-Shot Prompting LLMs with Limited Input}

Under identical limited-input conditions, few-shot prompting LLMs can exceed the 3/4-vote clinical threshold for the larger model but are generally below conventional supervised methods. Few-shot prompting with sufficient examples (\textit{k}=6-20) achieves F1-scores in the 0.59--0.88 range for DeepSeek, exceeding the clinical threshold (0.8000) in several configurations but typically trailing limited-input supervised baselines (FNN: 0.76--0.87; NCA-EVA: 0.69--0.82; Table~\ref{tab:exp5_conventional}) and full-feature LLM results (Table~\ref{tab:exp5}).

The zero-shot performance reveals important limitations: with limited input, 0-shot F1 ranges from 0.47 to 0.57 for SynthSeg+, far below the clinical threshold (0.8000), indicating that LLMs lack sufficient pre-trained knowledge about neuroimaging biomarker patterns without in-context examples. This finding contrasts with the full-feature setting (Figure~\ref{fig:exp4_f1}), where 0-shot achieves F1=0.74-0.76, suggesting that demographics and comprehensive ROI sets enable better zero-shot reasoning through richer context.

\begin{figure}[htbp]
\centering
\includegraphics[width=0.9\linewidth]{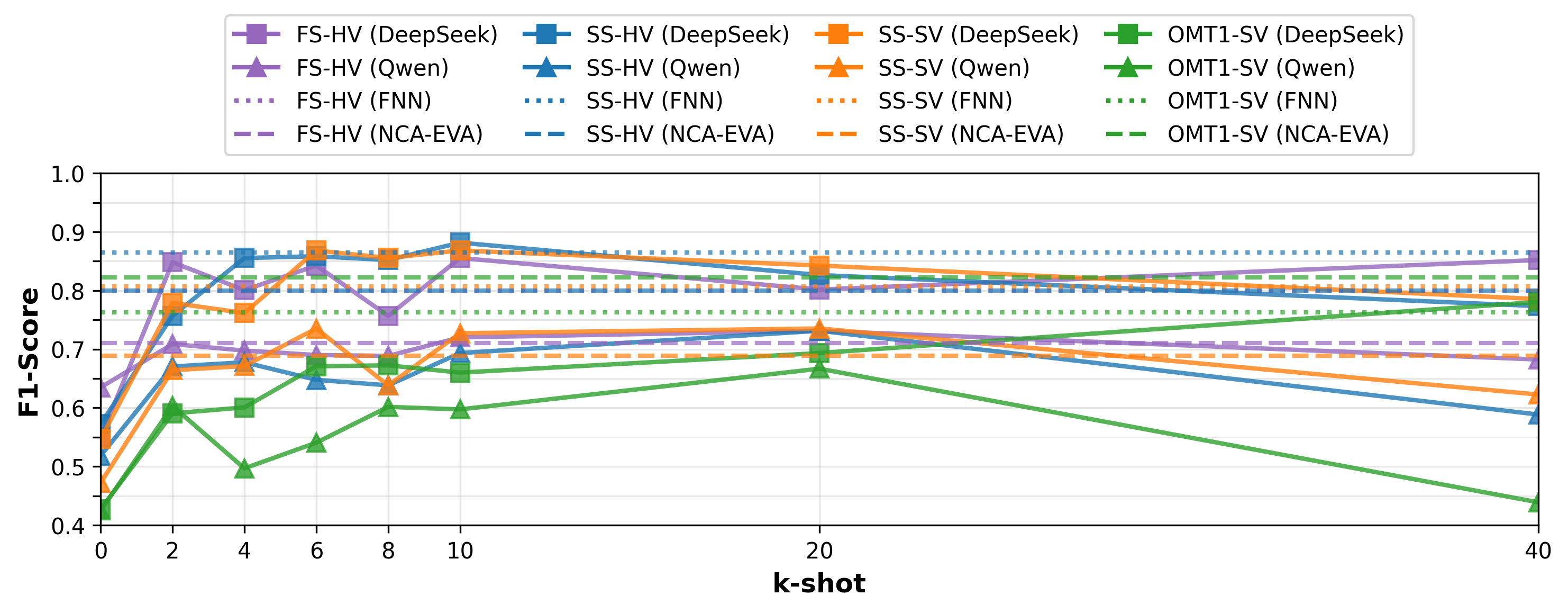}
\caption{\textbf{Limited-input performance comparison.} Few-shot LLM trajectories (square=DS, triangle=Qw) vs. clinical threshold baselines using identical 4-ROI input. Dotted lines: threshold F1. F1 Bootstrap 95\% CIs shown.}
\label{fig:exp5_f1_comparison}
\end{figure}

Comparing sample-efficiency trajectories between the full-feature curves (Table~\ref{tab:exp4}, Figure~\ref{fig:exp4_f1}) and the limited-input curves (Table~\ref{tab:exp5}, Figure~\ref{fig:exp5_f1_comparison}) shows that LLMs experience a larger performance drop than conventional supervised methods when features are restricted. This indicates that LLM reasoning benefits more from richer feature context or more in-context examples. At low \textit{k} (2--4), limited input causes pronounced degradation for LLMs, while supervised methods remain strong with the same 4 ROIs. With higher \textit{k} ($\geq$6), LLMs partially recover and can approach supervised levels but generally remain below FNN in this regime. Overall, with only 4 ROIs (8 features), supervised methods are preferable; LLMs require either more features or additional examples to close the gap.

Parcellation methods show fundamental differences in their information requirements, demonstrating a granularity-efficiency trade-off. OpenMAP-T1 consistently degrades with limited input across all \textit{k} values (7-44\% F1 loss vs. full input), indicating that its 280-region JHU atlas parcellation requires richer ROI information even with many examples. Finer parcellation granularity creates dependencies across a larger anatomical context: when only 4 critical ROIs are provided, the model cannot effectively reason about the relationships between these regions and the broader 280-region anatomical structure. In contrast, SynthSeg+'s coarser 100-region parcellation enables effective reasoning with minimal ROI sets when $\textit{k}\geq6$, demonstrating superior efficiency for data-scarce scenarios. The coarser granularity reduces the information gap between the 4 provided ROIs and the full parcellation space, allowing LLMs to more effectively leverage the critical anatomical information even when the full feature set is unavailable. Parcellation granularity should be matched to the available input features: finer parcellations excel when full feature sets are available, while coarser parcellations are more robust to feature subset limitations.

\begin{table}[htbp]
\centering
\caption{\textbf{Few-shot prompting LLM performance with limited input (4 ROIs only).} Acc=Accuracy, AUC=Area Under ROC Curve. F1 95\% BCa CIs via stratified bootstrap.}
\label{tab:exp5}
\scriptsize
\begin{tabular}{llcccc}
\toprule
\multirow{16}{*}{FS-HV} & \multirow{8}{*}{DeepSeek-v3} & 0 & 0.6771 & 0.4941 & 0.5500 [0.4444, 0.6850] \\
 &  & 2 & 0.8357 & 0.8573 & 0.8482 [0.7503, 0.9209] \\
 &  & 4 & 0.7753 & 0.8249 & 0.8007 [0.7013, 0.8703] \\
 &  & 6 & 0.8206 & 0.8722 & 0.8426 [0.7452, 0.8976] \\
 &  & 8 & 0.7453 & 0.8135 & 0.7568 [0.6389, 0.8352] \\
 &  & 10 & 0.8357 & 0.8694 & 0.8554 [0.7583, 0.9112] \\
 &  & 20 & 0.7757 & 0.8271 & 0.8017 [0.7029, 0.8786] \\
 &  & 40 & 0.8357 & 0.8452 & 0.8520 [0.7571, 0.9224] \\
\cmidrule{2-6}
 & \multirow{8}{*}{Qwen3-8B} & 0 & 0.6771 & 0.6507 & 0.6348 [0.5179, 0.7467] \\
 &  & 2 & 0.6771 & 0.7999 & 0.7088 [0.6055, 0.7851] \\
 &  & 4 & 0.6722 & 0.7851 & 0.6976 [0.6017, 0.7854] \\
 &  & 6 & 0.6490 & 0.7676 & 0.6901 [0.5975, 0.7726] \\
 &  & 8 & 0.6771 & 0.6619 & 0.6885 [0.5752, 0.7854] \\
 &  & 10 & 0.6949 & 0.7360 & 0.7201 [0.6195, 0.7970] \\
 &  & 20 & 0.6955 & 0.7520 & 0.7314 [0.6310, 0.8140] \\
 &  & 40 & 0.6490 & 0.6787 & 0.6824 [0.5724, 0.7660] \\
\midrule
\multirow{16}{*}{SS-HV} & \multirow{8}{*}{DeepSeek-v3} & 0 & 0.6667 & 0.5714 & 0.5726 [0.4000, 0.7347] \\
 &  & 2 & 0.7901 & 0.8520 & 0.7568 [0.6363, 0.8936] \\
 &  & 4 & 0.8765 & 0.9516 & 0.8554 [0.7500, 0.9545] \\
 &  & 6 & 0.8765 & 0.9488 & 0.8584 [0.7442, 0.9545] \\
 &  & 8 & 0.8765 & 0.9056 & 0.8520 [0.7554, 0.9545] \\
 &  & 10 & 0.9012 & 0.9294 & 0.8816 [0.7727, 0.9767] \\
 &  & 20 & 0.8519 & 0.8730 & 0.8264 [0.7179, 0.9333] \\
 &  & 40 & 0.8025 & 0.8452 & 0.7734 [0.6667, 0.9091] \\
\cmidrule{2-6}
 & \multirow{8}{*}{Qwen3-8B} & 0 & 0.5679 & 0.5464 & 0.5183 [0.3333, 0.6818] \\
 &  & 2 & 0.6790 & 0.8901 & 0.6705 [0.5854, 0.8636] \\
 &  & 4 & 0.7037 & 0.7298 & 0.6777 [0.5778, 0.8571] \\
 &  & 6 & 0.6543 & 0.7980 & 0.6478 [0.6000, 0.8782] \\
 &  & 8 & 0.6543 & 0.7714 & 0.6384 [0.5581, 0.8500] \\
 &  & 10 & 0.7160 & 0.7766 & 0.6934 [0.5852, 0.8571] \\
 &  & 20 & 0.7531 & 0.7980 & 0.7314 [0.6400, 0.8889] \\
 &  & 40 & 0.5926 & 0.7095 & 0.5886 [0.5555, 0.8372] \\
\midrule
\multirow{16}{*}{SS-SV} & \multirow{8}{*}{DeepSeek-v3} & 0 & 0.6049 & 0.5460 & 0.5469 [0.3749, 0.7273] \\
 &  & 2 & 0.8088 & 0.8272 & 0.7789 [0.6511, 0.9091] \\
 &  & 4 & 0.7901 & 0.8567 & 0.7616 [0.6500, 0.8936] \\
 &  & 6 & 0.8889 & 0.9194 & 0.8683 [0.7826, 0.9767] \\
 &  & 8 & 0.8765 & 0.9456 & 0.8554 [0.7619, 0.9545] \\
 &  & 10 & 0.8889 & 0.9361 & 0.8683 [0.7907, 0.9756] \\
 &  & 20 & 0.8642 & 0.9087 & 0.8426 [0.7619, 0.9545] \\
 &  & 40 & 0.8148 & 0.8484 & 0.7854 [0.6500, 0.9091] \\
\cmidrule{2-6}
 & \multirow{8}{*}{Qwen3-8B} & 0 & 0.5309 & 0.4698 & 0.4722 [0.2702, 0.6341] \\
 &  & 2 & 0.6790 & 0.8123 & 0.6642 [0.5854, 0.8571] \\
 &  & 4 & 0.6914 & 0.7782 & 0.6713 [0.5854, 0.8512] \\
 &  & 6 & 0.7531 & 0.8504 & 0.7353 [0.6363, 0.8889] \\
 &  & 8 & 0.6543 & 0.7298 & 0.6383 [0.5417, 0.8372] \\
 &  & 10 & 0.7407 & 0.7936 & 0.7272 [0.6511, 0.9000] \\
 &  & 20 & 0.7531 & 0.8504 & 0.7353 [0.6341, 0.8936] \\
 &  & 40 & 0.6296 & 0.7357 & 0.6227 [0.5500, 0.8333] \\
\midrule
\multirow{16}{*}{OMT1-SV} & \multirow{8}{*}{DeepSeek-v3} & 0 & 0.4444 & 0.4464 & 0.4273 [0.2857, 0.6512] \\
 &  & 2 & 0.6296 & 0.6163 & 0.5906 [0.4348, 0.7501] \\
 &  & 4 & 0.6420 & 0.6317 & 0.6008 [0.4500, 0.7620] \\
 &  & 6 & 0.7160 & 0.7095 & 0.6709 [0.5263, 0.8182] \\
 &  & 8 & 0.7037 & 0.7437 & 0.6725 [0.5455, 0.8262] \\
 &  & 10 & 0.7037 & 0.7063 & 0.6601 [0.5357, 0.8182] \\
 &  & 20 & 0.7160 & 0.7631 & 0.6934 [0.5882, 0.8571] \\
 &  & 40 & 0.8148 & 0.8147 & 0.7806 [0.6521, 0.9048] \\
\cmidrule{2-6}
 & \multirow{8}{*}{Qwen3-8B} & 0 & 0.4321 & 0.4603 & 0.4250 [0.2999, 0.6522] \\
 &  & 2 & 0.6216 & 0.5741 & 0.6031 [0.4889, 0.8002] \\
 &  & 4 & 0.5062 & 0.5817 & 0.4969 [0.3810, 0.7369] \\
 &  & 6 & 0.5432 & 0.6897 & 0.5407 [0.5000, 0.8000] \\
 &  & 8 & 0.6173 & 0.7079 & 0.6017 [0.4889, 0.7907] \\
 &  & 10 & 0.6173 & 0.6778 & 0.5974 [0.4865, 0.8000] \\
 &  & 20 & 0.6914 & 0.7246 & 0.6668 [0.5499, 0.8372] \\
 &  & 40 & 0.4444 & 0.5555 & 0.4390 [0.3500, 0.6957] \\

\bottomrule
\end{tabular}
\end{table}

Few-shot prompting with strategic example selection can surpass supervised methods even when both use identical limited inputs, particularly for larger models (DeepSeek achieves F1=0.72-0.84 with limited input, while smaller Qwen achieves F1=0.61-0.69). Larger foundation models with few-shot prompting can effectively leverage minimal critical anatomical information without requiring demographic covariates.
\section{Discussion}
\label{sec:discussion}

Our study evaluates existing methods across three experimental factors: \textbf{(1) parcellation method} (FS-HV, SynthSeg+ with 100-region atlas, OpenMAP-T1 with 280-region atlas), \textbf{(2) volumetry strategy} (hard volumetry via deterministic argmax vs. soft volumetry via probabilistic weighted summation), and \textbf{(3) classifier type} (clinical threshold classifiers using multi-ROI voting, feedforward neural networks, ensemble methods like NCA-EVA, and foundation models with binary vs. probability outputs and few-shot prompting k=0-40). This factorial design systematically evaluates how these factors interact to affect downstream AD classification performance, revealing generalizable patterns rather than isolated peak cases.


\subsection{Addressing Research Questions Through Three-Factor Analysis}

\subsubsection{RQ1: Parcellation Method Comparison}

SynthSeg+ and OpenMAP-T1 achieve numerically comparable or higher F1-scores than FS-HV across supervised and few-shot prompting settings while offering 45-150$\times$ faster inference. However, this pattern has important exceptions that reveal the importance of classifier-parcellation compatibility.

\paragraph{Classifier-Dependent Parcellation Performance.}
Under clinical thresholds calibrated to FS-HV atlas, FS-HV demonstrates superior performance (F1=0.72-0.87 across voting thresholds) compared to SynthSeg+ methods (F1=0.52-0.78), reflecting atlas-threshold compatibility. The inverted U-shape pattern for FS-HV (optimal at 2-3 votes) versus monotonic decline for SynthSeg+ (optimal at 1-vote) shows that threshold criteria optimized for FS-HV's volume distribution systematically misclassify SynthSeg+ volumes. FS-HV and SynthSeg+ produce systematically different absolute volume estimates due to distinct segmentation algorithms (surface-based morphometry vs. deep learning probabilistic segmentation), boundary definitions, and partial volume handling. When clinical thresholds calibrated for FS-HV volumes are transferred to SynthSeg+ volumes, performance degrades regardless of volumetry strategy, indicating that the fundamental issue is atlas-threshold compatibility rather than volumetry strategy per se. Clinical thresholds work best with their calibration source (FS-HV), while fast deep learning parcellation methods require recalibrated thresholds for optimal performance.

For supervised classifiers (FNN, NCA-EVA), both SynthSeg+ and OpenMAP-T1 achieve numerically comparable or higher F1-scores than FS-HV across configurations. Supervised performance falls within a narrow band with substantial CI overlap, indicating that fast deep learning parcellation methods can serve as practical alternatives to FS-HV for downstream classification.

For few-shot prompting LLMs, both SynthSeg+ and OpenMAP-T1 achieve comparable performance to FS-HV across \textit{k}-shot configurations. Fast deep learning parcellation methods (SynthSeg+: 12 sec/scan, OpenMAP-T1: 40 sec/scan) can serve as practical alternatives to FS-HV (30 min/scan) while maintaining competitive performance.

\paragraph{Parcellation Granularity and Sample Efficiency.}
Contrary to expectations that finer parcellation would yield systematically higher F1-scores, the performance curves reveal a granularity-efficiency trade-off with important implications for deployment. OpenMAP-T1's finer parcellation (280 regions) does not consistently outperform SynthSeg+ (100 regions), despite providing 2.8$\times$ more anatomical detail. For example, with DeepSeek, the OMT1-SV curve shows a gradual, steady increase with \textit{k}, requiring \textit{k}=6 to reach peak performance (F1=0.8714), while the SS-SV curve exhibits a sharp spike at \textit{k}=2 (F1=0.8684) before stabilizing. This 3$\times$ difference in required examples (2 vs. 6) with only a 0.003 F1 difference shows that coarser parcellations provide sufficient anatomical resolution for AD classification with substantially fewer examples. The limited-input experiments further illuminate this pattern: SynthSeg+'s coarser parcellation enables effective reasoning with minimal ROI sets when \textit{k}$\geq$6, while OpenMAP-T1 consistently degrades with limited input across all \textit{k} values, indicating that finer granularity creates dependencies across a larger anatomical context that cannot be compensated with feature subset limitations. Parcellation granularity should be matched to the available input features and annotation resources: finer parcellations excel when full feature sets are available and many examples can be provided, while coarser parcellations are more robust to feature subset limitations and data-scarce scenarios, making them preferable for resource-constrained deployments.

\subsubsection{RQ2: Classifier Paradigm Comparison}

Classifier paradigms show distinct performance patterns that depend on both parcellation method and volumetry strategy. Supervised methods (FNN, NCA-EVA) and few-shot prompting LLMs achieve comparable or superior performance to clinical thresholds, with each paradigm offering different trade-offs.

\paragraph{Supervised vs. Clinical Thresholds.}
Clinical thresholds work best with their calibration source (FS-HV) and degrade when transferred to other parcellations. Supervised classifiers achieve higher and more consistent performance across parcellation methods when training data are available, offering better generalization than fixed threshold rules.

\paragraph{FNN vs. NCA-EVA.}
Both FNN and NCA-EVA achieve competitive performance across parcellation methods. Patterns are mixed and depend on the parcellation--volumetry combination, with substantial CI overlap indicating comparable supervised performance overall.

\paragraph{Few-Shot Prompting vs. Supervised Methods.}
Few-shot prompting achieves competitive performance with supervised methods while requiring minimal labeled examples. The performance curves in Figure~\ref{fig:exp4_f1} illustrate several key trends. First, nearly all LLM configurations show a substantial F1 improvement from \textit{k}=0 to \textit{k}=2, demonstrating the power of in-context learning with just two examples. Second, performance for most methods begins to plateau after \textit{k}=6, suggesting diminishing returns for additional examples in this range. The curves show that few-shot LLMs with \textit{k}=2-6 examples consistently reach or exceed the performance of fully supervised FNN and NCA-EVA baselines. For example, the OMT1-SV+DeepSeek curve shows a steady upward trend, crossing the supervised baseline around \textit{k}=6, while SS-SV+DeepSeek spikes at \textit{k}=2 before stabilizing. This supports few-shot prompting as a practical alternative when labels are scarce.

Comparing full-feature (Figure~\ref{fig:exp4_f1}, Table~\ref{tab:exp4}) and limited-input (Figure~\ref{fig:exp5_f1_comparison}, Table~\ref{tab:exp5}) trajectories highlights a key difference in sample efficiency. When restricted to 4 ROIs (8 features), LLMs exhibit a larger performance drop than conventional supervised methods, indicating greater reliance on either richer feature context or more in-context examples to reason effectively. In this minimal-input regime, supervised FNN/NCA-EVA retain strong performance (F1~0.76--0.87), whereas few-shot LLMs generally require higher \textit{k} to approach these levels and typically remain below FNN. This pattern suggests a practical division of labor: use supervised classifiers for minimal-feature deployments; reserve few-shot LLMs for settings with richer feature sets or where interpretability and rationale output are prioritized.

\paragraph{LLM Optimization Strategies.}
Probability outputs outperform binary across most configurations, with an average F1 improvement of +35.0\% (range -2.5\% to +78.6\%). BCa analysis shows that probability outputs shift both lower and upper confidence bounds upward in 5 of 6 configurations, providing stronger evidence of superior performance despite wider intervals. Probability outputs not only improve mean performance but also provide more robust performance guarantees across the uncertainty range. The gains are substantially larger for smaller models (Qwen: +66.5\% average) compared to larger models (DeepSeek: +3.5\% average), likely because smaller models have less reliable internal confidence calibration. Requesting explicit probability outputs forces the model to provide calibrated confidence estimates that improve threshold selection. The one exception (SS-SV+DeepSeek showing -2.5\% degradation) may reflect a specific interaction between volumetry strategy and output format for this particular configuration, though the pattern is not consistently replicated across other configurations.

Multi-trial stability evaluation (5 trials per configuration), visualized as F1-score box plots in Figure~\ref{fig:exp6_stability}, shows acceptable variance across different few-shot selections with important architectural differences. DeepSeek (685B) consistently shows tighter interquartile ranges and fewer outliers than Qwen (8B) across \textit{k}-values, with average F1 standard deviation of 0.0176 versus 0.0498 for Qwen. This substantial stability difference (2.8$\times$ lower variance for DeepSeek) is smaller than the 85$\times$ parameter difference, indicating that architectural factors beyond scale influence stability. DeepSeek maintains stable variance across \textit{k} values (std range: 0.0052-0.0326), showing consistent performance regardless of example selection. In contrast, Qwen exhibits \textit{k}-dependent variance, with highest variance at intermediate \textit{k} values (e.g., std=0.1081 for SS-SV+Qw \textit{k}=6). Smaller models benefit from both zero examples (relying on pre-trained knowledge) and many examples (establishing clear patterns), but struggle with intermediate example counts where the balance between pre-trained knowledge and in-context learning is ambiguous. Zero-shot variance shows fundamental architectural differences: DeepSeek shows lowest variance at \textit{k}=0 (mean std=0.0138, CI width=0.07), indicating it leverages pre-trained knowledge more consistently, while Qwen shows higher zero-shot variance (mean CI width=0.16), indicating that smaller models require in-context examples to reduce both variability and uncertainty. Most configurations exhibit coefficient of variation (CV) $<$10\%, supporting deployment feasibility, though practitioners should expect $\pm$3-8\% F1 variation depending on example choice, particularly for smaller models at intermediate \textit{k} values.

Beyond accuracy, few-shot LLMs offer qualitative advantages that traditional classifiers do not. With simple prompt optimization (task framing, representative examples, and requesting probabilities), LLMs quickly close the initial zero-shot gap and approach supervised baselines while producing structured, case-specific rationales. These rationales make the model’s decision path explicit, can be steered to cite ROI-level evidence (e.g., hippocampal atrophy plus ventricular enlargement), and are useful for auditing borderline cases or guiding human review. This combination of competitive performance (by \textit{k}=2--6) and interpretable reasoning signals strong practical potential for AD detection when labels are scarce or transparency is required. Future work should standardize prompt design protocols, calibrate probability outputs across datasets, and evaluate how rationale quality relates to diagnostic utility in prospective settings.

\subsubsection{RQ3: Volumetry Strategy Effects}

The choice between hard and soft volumetry shows classifier-dependent impacts that reflect fundamental differences in how each approach processes information. Volumetry strategy has negligible impact for supervised classifiers but more variable effects for clinical thresholds and ensemble methods.

\paragraph{Volumetry Strategy with Clinical Thresholds.}
Hard and soft volumetry yield similar performance when applied to FS-HV volumes (the calibration source). When thresholds are transferred to SynthSeg+ volumes, performance degrades regardless of volumetry, indicating atlas--threshold compatibility as the main factor.

\paragraph{Volumetry Strategy with Supervised Classifiers.}
For supervised classifiers (FNN and few-shot LLMs), volumetry strategy shows negligible impact, indicating robustness to input representation. This robustness stems from the adaptive nature of supervised learning: FNNs and LLMs can learn feature transformations that map either discrete (hard volumetry) or continuous (soft volumetry) volume representations to the same discriminative decision boundaries. The nearly identical F1-scores (e.g., FNN SS-HV: 0.8575 vs. SS-SV: 0.8440) and CI widths (SS-HV: 0.1778 vs. SS-SV: 0.1793) confirm that both volumetry strategies provide equivalent information for supervised architectures, supporting the use of either strategy based on computational or data availability constraints.

Ensemble methods (NCA-EVA) show more variable sensitivity to volumetry strategy, with modest degradation on SS-SV (FNN: 0.8440 vs. NCA-EVA: 0.8128, -3.7\% relative to FNN). Feature selection mechanisms in ensemble methods may interact differently with probabilistic versus discrete inputs. The NCA dimensionality reduction may preserve discriminative patterns differently when operating on hard (discrete) versus soft (probabilistic) volumetry, though the differences are modest rather than substantial, indicating that volumetry choice remains secondary to classifier architecture selection.

\subsection{Interaction Patterns Across Factors}

\textbf{(1) Parcellation $\times$ Classifier:} The interaction between parcellation method and classifier type is the strongest determinant of performance. Clinical thresholds exhibit strong dependency on their calibration source (FS-HV), with performance degrading systematically when transferred to SynthSeg+ volumes due to systematic volume distribution differences. In contrast, supervised classifiers (FNN, NCA-EVA) and few-shot prompting LLMs demonstrate robustness to parcellation method, achieving comparable performance across FS-HV, SynthSeg+, and OpenMAP-T1. Adaptive learning (supervised or few-shot) can compensate for parcellation-specific volume differences, while fixed thresholds cannot. When using clinical thresholds, parcellation method choice is constrained to the calibration source; when using adaptive classifiers, parcellation method can be selected based on computational efficiency alone.

\textbf{(2) Volumetry $\times$ Classifier:} Volumetry strategy sensitivity depends on classifier architecture. Supervised methods (FNN, few-shot LLMs) show negligible sensitivity ($<$1\% F1 difference), indicating that adaptive learning can map either discrete or continuous volume representations to equivalent decision boundaries. Ensemble methods (NCA-EVA) show modest sensitivity (-3.7\% on SS-SV relative to FNN), suggesting that feature selection mechanisms interact differently with probabilistic versus discrete inputs. Volumetry choice is largely inconsequential for most supervised architectures but may warrant consideration for ensemble methods with feature selection.

\textbf{(3) Parcellation $\times$ Sample Efficiency:} The interaction between parcellation granularity and few-shot example requirements shows a granularity-efficiency trade-off. SynthSeg+'s coarser 100-region parcellation reaches competitive performance with fewer examples (\textit{k}=2-4), while OpenMAP-T1's finer 280-region parcellation requires more examples (\textit{k}=6-10) to achieve comparable performance. Coarser parcellations provide sufficient anatomical context for AD classification with minimal examples, while finer parcellations require more examples for the model to learn effective reasoning patterns across the richer anatomical structure. The limited-input experiments show that coarser parcellations are more robust to feature subset limitations, while finer parcellations degrade more substantially when only 4 critical ROIs are provided.

These interaction patterns show that performance optimization requires holistic pipeline design rather than component-level optimization. The strong parcellation-classifier interaction indicates that deployment constraints (computational efficiency, data availability, annotation resources) should guide both parcellation and classifier selection simultaneously.

\subsection{Evidence-Based Selection Guidelines}

\textbf{Clinical Deployment with Threshold Rules:} Prefer FS-HV (calibration source) for thresholds; if runtime is prohibitive, SynthSeg+ provides acceptable screening performance at 150$\times$ faster inference. \textbf{Research Studies with Labeled Training Data:} Both FNN and NCA-EVA with SynthSeg+ or OpenMAP-T1 achieve competitive performance; choose based on compute and maintenance constraints. Few-shot prompting reaches similar performance with \textit{k}=2--6 examples when labels are scarce. \textbf{Developing New Parcellation Methods:} Evaluate full pipelines and sample-efficiency trends across \textit{k}, rather than single operating points that may overfit; align granularity and volumetry with the downstream task and data regime.

\subsection{Limitations and Broader Implications}

Several limitations warrant future investigation. OASIS-1's homogeneity (single scanner, protocol) limits generalization assessment; multi-site validation (ADNI, OASIS-3, UK Biobank) is essential. Age and sex imbalances may bias results despite mitigation strategies. Clinical practice combines multiple biomarkers; our findings represent volumetric-only upper bounds. A single 60/20/20 split limits robustness; k-fold cross-validation would strengthen generalizability. Soft volumetry testing was limited to shallow FNNs; spatial models that preserve probabilistic boundary information may demonstrate advantages. We compare integrated pipelines rather than isolated parcellation quality. Parameter-efficient fine-tuning (LoRA) may improve beyond few-shot prompting. Local deployment requires substantial GPU resources; API alternatives raise privacy concerns. Few-shot prompting's 2--6 example requirement democratizes medical AI for rare diseases; task-aware parcellation optimization represents promising directions.

\section{Conclusion}
\label{sec:conclusion}

This study decouples brain parcellation from classification and systematically benchmarks fast deep learning parcellation methods (SynthSeg+, OpenMAP-T1) against the FreeSurfer (FS) clinical baseline through downstream AD classification on OASIS-1. Our factorial design evaluates three parcellation methods, two volumetry strategies (hard vs. soft), and four classifier paradigms (clinical thresholds, supervised feedforward networks, ensemble methods, and foundation models with zero/few-shot prompting), with all results quantified using BCa Bootstrap 95\% confidence intervals. Given the modest test size (n=81), we emphasize effect sizes and consistent directional patterns rather than formal significance testing.

\textbf{Key findings:} (1) Fast parcellations achieve F1-scores (0.84-0.87) comparable to FS-HV (0.72-0.87) with 45-150$\times$ speedups when paired with adaptive classifiers, but clinical thresholds calibrated for FS-HV do not transfer reliably to other parcellations---highlighting atlas--classifier compatibility as a critical constraint for rule-based workflows. (2) Volumetry strategy (hard vs. soft) has negligible impact for supervised classifiers ($<$1\% F1 difference), though ensemble methods show modest sensitivity. (3) Few-shot prompting exhibits rapid 0$\rightarrow$2 shot improvement and saturates around \textit{k}\,$\approx$\,6, reaching supervised-level performance (F1=0.74-0.87) with minimal examples. Probability outputs consistently improve F1 (mean +35.0\%) by shifting both BCa CI bounds upward in 5 of 6 configurations, especially for smaller models (Qwen: +66.5\% vs. DeepSeek: +3.5\%). Multi-trial evaluation confirms acceptable LLM stability (mean CV $<$10\%), with larger models more stable. (4) Parcellation granularity shows a sample-efficiency trade-off: coarser parcellations (100 regions) reach competitive performance with fewer examples than finer parcellations (280 regions), with minimal F1 difference. (5) Under limited-input conditions (4 ROIs), conventional supervised methods retain strong performance (FNN: F1=0.76-0.87; NCA-EVA: F1=0.69-0.82), whereas few-shot LLMs require richer feature context or more examples to close the gap, indicating that supervised methods are preferable for minimal-feature deployments.

\textbf{Practical guidance:} For rule-based workflows, use FS-HV with its calibrated clinical thresholds. When labeled data are available, fast parcellations paired with supervised classifiers (FNN: F1=0.83-0.86; NCA-EVA: F1=0.81-0.86) offer high performance and efficiency. In data-scarce scenarios, few-shot LLMs with probability outputs and \textit{k}=2-6 examples provide a competitive alternative, offering interpretable rationales and calibrated probabilities for auditing borderline cases. Volumetry choice is largely inconsequential for supervised methods. Beyond accuracy, few-shot LLMs bring workflow benefits: with modest prompt optimization they approach supervised baselines while providing structured case rationales that can support audit and triage in borderline cases. Future work should formalize prompt design guidelines, evaluate rationale quality in prospective settings, and study calibration transfer across datasets.

To our knowledge, this is the first systematic study that decouples parcellation from classification and benchmarks fast parcellations, volumetry strategies, and LLM-based classifiers against FS-HV baselines with bootstrap uncertainty quantification on OASIS-1.

\section*{Ethics \& Reproducibility}
\textbf{Ethics:} All experiments use publicly available OASIS-1 data with appropriate data use agreements. No new human subjects data was collected. 

\textbf{Reproducibility:} Code, trained models, evaluation protocols, and data splits are available from the corresponding author upon request and will be made publicly available at a project repository upon publication.

\bibliography{references}

\end{document}